\documentclass[useAMS,usenatbib]{mn2e} 
\usepackage{amsmath}
\usepackage{amssymb}
\usepackage[dvips]{graphicx}
\usepackage{epsfig} 
\title[Weather at Sierra Negra]{Weather at Sierra Negra: 7.3-year statistics
  and a new method to estimate the temporal fraction of cloud cover}

\author[E. Carrasco et al.]
{
E. Carrasco,$^{1}$\thanks{E-mail: bec@inaoep.mx}
A. Carrami{\~n}ana,$^{1}$
R. Avila,$^{2}$
C. Guti{\'e}rrez,$^{1}$
J.~L. Avil{\'e}s,$^{1,2}$
\newauthor
J. Reyes,$^{1}$
J. Meza$^{1}$
and O. Yam$^{3}$\\
$^{1}$Instituto Nacional de Astrof{\'\i}sica, \'Optica y Electr\'onica, 
Luis Enrique Erro 1, Tonantzintla, Puebla, C.P. 72840,  M\'exico\\
$^{2}$Centro de Radioastronom{\'\i}a y Astrof{\'\i}sica, UNAM, Apartado
Postal 3-72, Morelia, Michoac\'an, C.P. 58089, M\'exico\\
$^{3}$Universidad  de Quintana Roo, Boulevard Bah\'{\i}a S/N, Chetumal
77019, Quintana Roo, M\'exico 
}

\begin{document}

\date{}

\pagerange{\pageref{firstpage}--\pageref{lastpage}} \pubyear{2008}

\maketitle

\label{firstpage}

\begin{abstract}
Sierra Negra,  one of the highest peaks in central Mexico, is the site of 
the Large Millimeter Telescope.  We describe the first results of a 
comprehensive analysis of the weather data measured in situ from October 
2000 to February 2008 to be used as a reference  for future activity  
in the site.  We compare  the data from two different stations at the summit
considering the accuracy of both instruments.  We analysed  the diurnal, 
seasonal  and annual cycles for all the  parameters.  
The thermal stability is remarkably good, crucial for a good performance of the 
telescopes. From the  solar radiation data we  developed a new  method
to estimate the fraction of time when the sky is clear of clouds. 
We show that our measurements are consistent with a warm standard
atmosphere model.  The conditions at the site are benign and stable
given its altitude,  showing that Sierra Negra is a extremely good  site
for millimeter  and high energy   observations.   

\end{abstract}

\begin{keywords}
site testing --- atmospheric effects
\end{keywords}

\section{Introduction}
High altitude astronomical sites are a scarce commodity with increasing demand. 
A thin atmosphere can make a substantial difference in the performance of 
scientific research instruments like millimeter-wave telescopes or water 
\v{C}erenkov observatories. In our planet reaching above 4000 metres involves 
confronting highly adverse meteorological conditions. Sierra Negra, the site 
of The Large Millimeter Telescope/El Gran Telescopio Milim\'etrico (LMT) is 
exceptional in being one of the highests astronomical sites available with 
endurable weather conditions. The LMT site combines high altitude (4580~m) and 
low atmospheric water content. The water vapor opacity has been monitored 
since 1997 with radiometers working at 225 GHz showing that the  zenith 
transmission at the site is better than 0.89 at 1 mm during 7 months of the 
year and better than 0.80 at 850 microns during 3 months of the year \citep{Hughes08}.
There is no telescope as massive as the LMT above 4500~metres anywhere else 
and one can barely expect to operate at that altitude with temperatures above 
freezing. The development of the LMT site led to the interest and development 
of other scientific facilities benefiting from the high altitude conditions 
and sharing the same basic infrastructure. In July 2007 the base of Sierra 
Negra was selected as the site for the High Altitude Water \v{C}erenkov 
(HAWC) gamma-ray observatory, an instrument whose performance depends 
critically on its 4100~m altitude location.

\begin{figure*}
\includegraphics[width=\columnwidth]{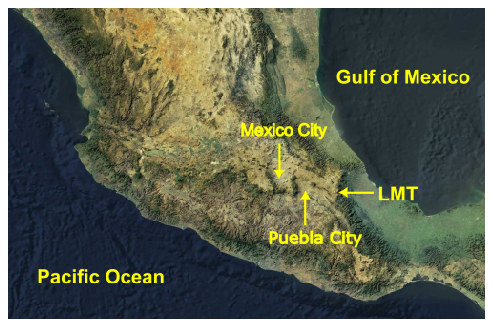}
\includegraphics[width=0.87\columnwidth]{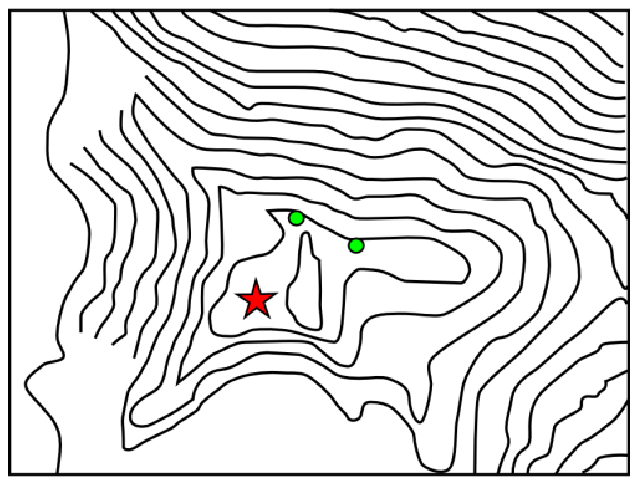} 
\caption{Left: map of Mexico indicating the location of Sierra Negra site [from Conabio site: www.conabio.gob.mx].
 Right: site zoom [from INEGI site: www.inegi.org.mx]. The star on the left low corner is the LMT, 
the open circles show  the positions of the meteorological stations. The distance between the stations is 110~m.  See the electronic edition of MNRAS 
for a color version of this figure. }
\label{zoom}
\end{figure*}

\section{The Sierra Negra site}

Sierra Negra, also known as Tliltepetl, is a 4580~meter volcano inside the 
Parque Nacional Pico de Orizaba, a  national park named after the highest 
mountain of Mexico.  With an altitude of 5610~m\footnote{Instituto 
Nacional de Estad\'{i}stica, Geograf\'{i}a e Inform\'atica (INEGI)  
official figure.} Pico de Orizaba, also known as Citlaltepetl, is one
of the seven most prominent peaks in the world,  where 
prominent   is related with the dominance  of the mountain over the
region\footnote{Topographic prominence is defined as the elevation 
difference between the peak summit and the lowest
contour level that encircles that summit but does not encircle any higher
summit.} \citep{Press82}. The Parque National  Pico de Orizaba has an area 
of 192~km$^{2}$ enclosing the two volcanic peaks,  separated by only 7~km 
from top to top, and their wide bases. Tliltepetl is 
an inactive volcanic cone formed 460,000 years ago, much earlier than 
Citlaltepetl whose present  crater was created just 4100 years ago and has 
a record of activity within  the last 450  years, including the flow of 
0.1~km$^{3}$ of lava in 1566 and a last eruptive event in 1846 
\citep{Hoskuldsson93, Rossotti05}. 
These two peaks are located at the edge of the Mexican plateau which  drops at the 
East to reach the Gulf of Mexico at about 100~km distance, as shown on
the left side of  Fig.~\ref{zoom}. The weather  of the site is influenced
by the dry weather of the high altitude central Mexican plateau and humid 
conditions coming from the Gulf of Mexico \citep{Erasmus02}.

In February 1997 Sierra Negra was selected as the site of the LMT, a 50~m antenna 
for astronomical observations in the 0.8 - 3  millimeter range. The top of Sierra 
Negra, defined now by the position of the telescope, on the right side of
Fig.~\ref{zoom}, has Universal Transverse Mercator (UTM) and geographical coordinates 
 $\{ x_{utm} = 677 450, y_{utm} = 2 100 092\}$ and $\{
  97^{\circ}\, 18^{\prime}\, 51.7^{\prime\prime}\, {\rm longitude\, West},\,
 18^{\circ}\,59^{\prime}\,08.4^{\prime\prime}\,\rm latitude\, North\}$ respectively.
The development of the  LMT site led to the installation of further scientific facilities 
benefiting from its strategic location and basic infrastructure like the 5~m radio telescope
 RT5, a solar neutron telescope and cosmic ray detectors, among others. 
In July 2007 the base of Sierra Negra, about 500~m below the summit, was chosen as the 
site of the High Altitude Water \v{C}erenkov (HAWC) observatory, a $\sim 20000\,\rm 
m^{2}$ water \v{C}erenkov observatory for mapping and surveying the high energy 
$\gamma$-ray sky. HAWC will be complemented by two atmospheric air \v{C}erenkov 
telescopes, the OMEGA (Observatorio MExicano de GAmmas) formerly part of the HEGRA 
array~\citep{hegra}.

The seeing of Sierra Negra was monitored between 2000 and 2003 to quantify the 
potential of the site for optical astronomy. The site has a median seeing of 0.7", 
consistent with of a prime astronomical site~\citep{Carrasco03}. The wind velocity 
at 200 mbar has been analyzed using the NOAA NCEP/NCAR reanalysis database 
showing that Sierra Negra is comparable to the best observatory sites as Mauna 
Kea in terms of applying adaptive optics techniques such as slow wavefront corrugation 
correction \citep{Carrasco05}, based on the premise that global circulation of 
atmospheric winds at high altitude can be used as a criterion to establish 
the suitability of a site for the development of adaptive optics technique as the
wind velocity at 200~mbar is strongly correlated to the average wavefront velocity allowing
to compute the  coherence time $\tau_{o}$  \citep{Sarazin02}. 

Different scientific facilities seek particular conditions and their dependence
on meteorological conditions vary. Among the Sierra Negra facilities we can note:
\begin{itemize}
\item the Large Millimeter Telescope requires minimum atmospheric opacity in
the millimeter range, which translates in a reduced water vapour  column density.
According to design specifications, LMT operation at 1~mm require wind velocities 
below 9~m~s$^{-1}$ and the antenna is able to survive winds up to 250 km/h (69.4~m~s$^{-1}$).
\item the RT5 5~m radio telescope will operate at 43 and 115~GHz for observations 
of the Sun. Nighttime work will focus on interstellar masers and monitoring of 
mm-wave bright active galactic nuclei. RT5 requires absence of clouds in the line
of sight.
\item optical and atmospheric \v{C}erenkov telescopes require clear nights
and relatively low humidity (below 80\%) during nighttime. 
\item water \v{C}erenkov observatories like HAWC seek high altitude environments 
which allow for a deep penetration of atmospheric particle cascades. They are 
basically immune to weather,
although freezing conditions and large daily temperature cycles are concerns.
The same applies to small cosmic ray detectors installed at Sierra Negra summit.
\end{itemize}

\begin{table}
\centering
\caption {Positions of the weather stations relative to the LMT; $x$
 increases to the East and $y$ to the North.} \label{positions}
   \begin{tabular}{lrrr}
   \hline
 & \multicolumn{3}{c}{Relative location} \\
Instrument  & x(m) & y(m) & z(m) \\
\hline
LMT     & 0 & 0 & 0 \\
Davis  & 139 & 65 & --15 \\
Texas & 40 & 105 & 0 \\
\hline
\end{tabular}
\end{table}

\section{Instrumentation and location}
The weather data presented here were acquired with three instruments:
\begin{enumerate}
\item {\bf a  Davis  meteorological station}, hereafter named ``Davis", located on 
a 5~m tower about (139, 65) metres (E,N) from the LMT position. Most of the data 
shown in this paper comes from this station. It has been operational since October 2000 up to now,
and is installed at the base of the former seeing monitor. The station tower is at the 
edge of a sharp slope facing North East just above the HAWC site -that
 is approximately 430m E, 1010m N and 500m below the summit. 
The Davis station consists of temperature and humidity sensors enclosed on a radiation 
shield, a barometre, an anenometer, a control console and a data logger:
\begin{itemize}
\item the temperature sensor is a platinum wire thermistor with a resolution of 
$0.1^{\circ}\rm C$ and a nominal accuracy of $\pm 0.5^{\circ}\rm C$. 
\item the relative humidity (RH) sensor is a film capacitor element providing a 
resolution of 1\% with an accuracy of $\pm 3\%$ for RH between 0 and 90\% and 
$\pm 4\%$ above 90\%. 
\item the barometre has a resolution of $0.1\,\rm mbar$ and an accuracy of 
$\pm 0.4\,\rm mbar$ in the measurement of atmospheric pressure.
\item the wind monitor consists of a three cup anemometre providing a resolution
of 0.4~m~s$^{-1}$  and  accuracy better than 0.9 m~s$^{-1}$   
  for a wind speed interval between  0.9 and 78 m~s$^{-1}$. \footnote{Specified by the manufacturer in  English units: wind speed resolution of 1~mile/hr and nominal accuracy better than 2~mile/hr for an 
interval between 2 and 175~mile/hr}
 \end{itemize}
\item {\bf a backup Davis station}, hereafter named ``Backup", of same model and
characteristics as  the main Davis station, was temporarily installed at the same 
position and operated from April 2002 until November 2003, when it ceased functioning.
\item {\bf a Texas Electronics weather station}, hereafter named ``Texas", consists of:
\begin{itemize}
\item a temperature sensor, of model TT-101QR, made of a linear thermistor resistor of 
0.1$^\circ$C resolution and 0.5$^\circ$C accuracy \footnote{Specified in  English units: 
the T accuracy is 1$^\circ$F} in the range  $-34^{\circ}\rm C$ to $ 43^{\circ}\rm C$;
\item a humidity sensor consisting of a thin film capacitor with a resolution of 1\%
and an accuracy of $\pm 3\% $. 
\item  a radiation sensor made of a solar panel inside a 
glass dome to obtain maximum cosine response to the Sun's radiation. The nominal 
range is up to 1400~W/m$^{2}$ with a resolution of 1~W/m$^{2}$ and $5\%$ accuracy.
\item a wind monitor consisting  of three anemometre cups providing a resolution
of $0.2\,\rm m\, s^{-1}$ and an accuracy better than 0.5~m~s$^{-1}$.
\end{itemize} 
The Texas weather station has also a barometre, but the readings of
atmospheric pressure were found to be spurious.
\end{enumerate}
The Texas station was installed 
at about (40, 105) metres (E,N) from the LMT and 110~metres apart from the Davis station
and  15~m higher than the Davis. Taking the LMT as the reference point the coordinates of both
stations are shown in table \ref{positions}.
The relative  locations of the LMT  and the weather stations are shown in 
Fig. \ref{zoom}

\begin{table*}
 \centering
  \caption{Data coverage of Davis weather 
station in percentage.\label{cov_davis_humedad}}
  \begin{tabular}{|l|rrrrrrrrr|r|}
  \hline
Month &  2000 & 2001 & 2002 & 2003 & 2004 & 2005 & 
2006 & 2007 & 2008 & All\\
  \hline
 January      &  &  65 &  87 &  95 &  65 &  99 &  95 &  77 &     99 &  85 \\
 February   &  &  69 &  92 &  99 &  68 &  92 &  77 &  19 &     60 &  76 \\
 March     &  &   0 &  99 &  86 &  57 &  70 &  79 &   0 &  &  56 \\
 April     &  &   0 &  85 &  91 &  96 &  87 &  93 &   0 &  &  64 \\
 May      &  &  63 &  99 &  74 &  89 &  78 &  79 &   0 & &  69 \\
 June     & &  56 &  94 &  99 &  83 &  88 &  78 &   0 &  &  71 \\
 July     & &  37 &  97 &  92 &  85 &  77 &  56 &  67 &  &  73 \\
 August     &  &  94 &  98 &  89 &  86 &  73 &  25 &  79 &  &  78 \\
 September &  &  91 &  93 &  55 &  76 &  98 &  38 &  94 & &  78 \\
 October    &      3 &  97 &  46 &  90 &  90 &  89 &  76 &  24 & &  73 \\
 November  &     14 &  75 &  44 &  81 &  77 &  60 &  60 &   0 &  &  51 \\
 December  &     37 &  82 &  78 &  66 &  48 &  36 &  90 &  63 &  &  63 \\
\hline
 Year total &     27 &  61 &  84 &  85 &  77 &  79 &  70 &  35 &     98 &  70\\
\hline
 Dry        &     26 &  48 &  81 &  86 &  68 &  74 &  82 &  27 &     98 &  66 \\
 Wet        &     54 &  73 &  88 &  83 &  85 &  84 &  59 &  44 &  &  73 \\
\end{tabular}
\end{table*}

\begin{table*}
 \centering
  \caption{Data coverage of relative humidity  from the Texas weather 
station in percentage.\label{cov_texas_humedad}}
  \begin{tabular}{|l|rrrrrrr|r|}
  \hline
Month & 2002 & 2003 & 2004 & 2005 & 2006 & 2007 & 2008 & All\\
  \hline 
January    &  &  88 &  89 &   0 &  82 &  64 &     27 &  58 \\
 February   &  &  93 &  88 &   0 &  83 &  80 &     91 &  73 \\
 March      &  & 100 &  75 &  68 &  93 &  87 &     41 &  77 \\
 April      &     48 &  84 &   0 &  61 &  95 &  69 &  &  59 \\
 May        &     77 &   0 &  40 &  38 &  78 &  10 &  &  40 \\
 June       &     88 &   0 &   0 &  84 &  71 &  23 &  &  44 \\
 July       &     97 &   0 &   0 & 100 & 100 &   8 &  &  50 \\
 August     &     77 &  61 &   0 &  89 &  42 &   0 &  &  45 \\
 September  &     85 &  51 &   0 &  99 & 100 &  41 &  &  63 \\
 October    &     99 &  73 &   0 & 100 &  85 &  12 &  &  61 \\
 November   &     94 &  64 &   0 &  87 &  59 &  68 &  &  62 \\
 December   &     98 &  69 &   0 &  74 &  80 &  56 &  &  63 \\
\hline
 Year total &     85 &  57 &  24 &  67 &  81 &  43 &     52 & 58 \\
\hline
 Dry        &     80 &  83 &  42 &  49 &  82 &  71 &     52 & 65 \\
 Wet        &     87 &  31 &   6 &  85 &  79 &  15 &      0 & 51 \\
\end{tabular}
\end{table*}

\begin{table*}
 \centering
  \caption{Solar radiation data coverage from the Texas weather 
station in percentage.\label{cov_texas_radiation}}
  \begin{tabular}{|l|rrrrrrr|r|}
  \hline
Month & 2002 & 2003 & 2004 & 2005 & 2006 & 2007 & 2008 & All\\
  \hline 
January  &      &   87  &  52 &   60  &  81 &   64 &   28  &  62 \\
 February &     &   93  &  87 &   69  &  83 &   80 &   91  &   84 \\
 March    &     &   99  &  57 &   98  &  93 &   88 &   40  &   79  \\
 April    &   17  &   83  &   8 &   61  &  93  &  72  &    &   56 \\
 May      &   39  &   0   &  25 &   37  &  82  &  10  &    &   32 \\
 June     &   87  &   0   &   0 &   84  &  71 &   46  &    &  48\\
 July     &   73  &   0   &  29 &   99  &  99 &   8   &    &   51 \\
 August   &   77  &  26   &  99 &   88  &  43 &   32  &    &   61 \\
 September &  84  &  51   &  50 &   99  &  99 &   59  &     &   74 \\
 October   &  99  &  74   &  44 &   99  &  85 &   11  &     &   69 \\
 November  &  93  &  64   &  45 &   86  &  66 &   69  &     &   71 \\
 December  &  97  &  68   &  46 &   74  &  81 &   56  &     &   70 \\
\hline
 Year       &   73 &   52 &   44 &  80  &  81 &   48  &  52 &  62\\ 
\hline
 Dry        &   68 &   83 &   49 &  75  &  83 &   72  &  52 &   70 \\
 Wet        &   76 &   24 &   41 &  84  &  80 &   27  &   0 &   55 \\
\end{tabular}
\end{table*}

\begin{table*}
\caption{Hourly data coverage of Davis weather station.\label{cov-davis-2}}
\begin{tabular}{lrrrrrrrrrrrrr}
\hline
{} & {Jan} &  {Feb} &  {Mar} &  {Apr} &  {May} &  {Jun} &  
                   {Jul} &  {Aug} & {Sep} &  {Oct} &  {Nov} &  {Dec} & {Per Hour} \\
\hline
   0h &  85 & 76 & 55 & 65 & 69 & 72 & 74 & 78 & 78 & 73 & 52 & 62 &  70 \\
   1h &  84 & 77 & 55 & 64 & 69 & 72 & 75 & 78 & 77 & 73 & 52 & 62 &  70 \\
   2h &  85 & 77 & 56 & 64 & 69 & 72 & 74 & 78 & 77 & 73 & 52 & 62 &  70 \\
   3h &  85 & 77 & 55 & 64 & 69 & 73 & 74 & 78 & 78 & 73 & 52 & 62 &  70 \\
   4h &  84 & 76 & 55 & 63 & 69 & 73 & 74 & 78 & 78 & 73 & 52 & 62 &  70 \\
   5h &  85 & 77 & 56 & 63 & 70 & 73 & 74 & 78 & 78 & 73 & 51 & 62 &  70 \\
   6h &  85 & 77 & 56 & 64 & 70 & 72 & 74 & 78 & 78 & 73 & 51 & 62 &  70 \\
   7h &  85 & 77 & 56 & 64 & 70 & 72 & 74 & 78 & 78 & 74 & 52 & 63 &  70 \\
   8h &  85 & 77 & 57 & 64 & 69 & 72 & 73 & 79 & 78 & 74 & 52 & 63 &  70 \\
   9h &  85 & 75 & 56 & 63 & 69 & 71 & 73 & 78 & 78 & 72 & 51 & 63 &  69 \\
  10h &  86 & 75 & 56 & 64 & 68 & 70 & 72 & 77 & 78 & 72 & 51 & 63 &  69 \\
  11h &  86 & 76 & 57 & 65 & 69 & 71 & 73 & 77 & 78 & 72 & 51 & 63 &  70 \\
  12h &  86 & 76 & 56 & 64 & 69 & 71 & 72 & 77 & 77 & 72 & 52 & 63 &  69 \\
  13h &  86 & 76 & 56 & 65 & 68 & 70 & 72 & 77 & 76 & 72 & 52 & 63 &  69 \\
  14h &  86 & 75 & 56 & 64 & 68 & 70 & 72 & 77 & 77 & 73 & 52 & 62 &  69 \\
  15h &  87 & 74 & 56 & 65 & 68 & 70 & 72 & 77 & 77 & 73 & 52 & 62 &  69 \\
  16h &  87 & 74 & 56 & 66 & 69 & 70 & 72 & 78 & 77 & 72 & 51 & 63 &  70 \\
  17h &  86 & 76 & 56 & 66 & 69 & 70 & 72 & 78 & 77 & 73 & 52 & 63 &  70 \\
  18h &  86 & 76 & 56 & 66 & 69 & 70 & 72 & 77 & 77 & 72 & 51 & 62 &  70 \\
  19h &  86 & 75 & 56 & 65 & 69 & 70 & 72 & 78 & 78 & 73 & 51 & 62 &  70 \\
  20h &  85 & 76 & 56 & 65 & 69 & 69 & 71 & 77 & 77 & 72 & 51 & 62 &  69 \\
  21h &  85 & 75 & 56 & 65 & 68 & 70 & 72 & 76 & 78 & 72 & 50 & 63 &  69 \\
  22h &  85 & 75 & 55 & 64 & 69 & 70 & 72 & 78 & 77 & 72 & 51 & 63 &  69 \\
  23h &  85 & 75 & 55 & 65 & 68 & 71 & 73 & 78 & 78 & 71 & 52 &  63 &  69 \\
\hline
{Per Month}  &  85 & 76 & 56 & 64 & 69 & 71 & 73 & 78 & 78 & 73 & 51 & 63 & {\bf 70} \\
\hline
\end{tabular}
\end{table*}

\begin{table*}
\caption{Hourly data coverage of humidity from the Texas weather station.
\label{cov-hum-1}}
\begin{tabular}{lrrrrrrrrrrrrr}
\hline
{} & {Jan} &  {Feb} &  {Mar} &  {Apr} &  {May} &  {Jun} &  
                   {Jul} &  {Aug} & {Sep} &  {Oct} &  {Nov} &  {Dec} & {Per Hour} \\
\hline
    0h &    59 &   67 &  77 &  56 & 40 &   44 &  51 &  45 &  63 &  61 &   60 &   63 &  58 \\
    1h &    59 &   67 &  77 &  56 & 39 &   42 &   51 & 45 & 63 & 60 &   60 &   62 &  57 \\
    2h &    59 &   67 &  77 &  56 & 39 &   42 &   51 & 45 & 63 & 60 &   60 &   62 &  57 \\
    3h &    58 &   67 &  77 &  56 & 39 &   42 &   51 & 44 & 63 & 59 &  60 &   62 &  57 \\
    4h &    58 &   67 &  77 &  56 & 39 &   42 &   51 & 44 & 63 & 59 &  60 &   62 &  57 \\
    5h &    57 &   67 &  77 &  56 & 39 &   42 &   50 & 43 & 63 & 59 &  60 &   62 &  56 \\
    6h &    57 &   67 &  77 &  55 & 39 &   42 &   49 & 42 &   62 & 59 &  59 &  62 &  56 \\
    7h &    56 &   67 &  77 &  56 & 40 &   43 & 49 & 43 & 62 & 59 &  59 &  62 &  56 \\
    8h &    57 &   67 &  77 &  60 & 41 &   43 & 49 & 45 & 62 & 59 &  61 &   62 &  57 \\
    9h &    57 &   67 &  78 &  61 & 42 &   43 & 49 & 45 & 63 & 59 &  62 & 62 &  57 \\
  10h &    57 &   67 &  78 &  61 & 41 &   44 & 50 & 45 & 63 & 59 &  62 & 62 &  58 \\
  11h &    57 &   67 &  77 &  61 & 41 &   44 & 51 & 44 & 62 & 60 &   63 & 62 &  58 \\
  12h &    58 &   68 &  77 &  61 & 40 &   44 & 51 & 45 & 62 & 62 & 64 & 62 &  58 \\
  13h &    59 &   69 &  76 &  62 & 41 &   44 & 51 & 46 & 63 & 64 & 65 & 63 &  59 \\
  14h &    59 &   70 &  77 &  62 & 42 &   45 & 51 & 46 & 63 & 64 & 64 & 64 &  59 \\
  15h &    60 &   70 &  77 &  62 & 42 &   46 & 51 & 45 & 63 &  64 & 64 & 64 &  59 \\
  16h &    60 &   69 &  78 &  63 & 42 &   46 & 51 & 45 & 62 &   64 & 65 & 64 &  59 \\
  17h &    60 &   69 &  78 &  63 & 42 &   46 & 51 & 45 & 62 &   65 & 65 & 65 &  59 \\
  18h &    60 &   68 &  78 &  63 & 42 &   46 & 51 & 46 & 63 &   64 & 65 & 65 &  59 \\
  19h &    60 &   68 &  77 &  63 & 41 &   46 & 51 & 46 & 63 &   64 & 64 & 64 &  59 \\
  20h &    60 &   68 &  77 &  63 & 40 &   46 & 51 & 46 & 62 &   64 & 64 & 63 &  59 \\
  21h &    60 &   69 &  77 &  60 & 40 &   46 & 51 & 45 & 62 &   63 &   63 &   63 &  59 \\
  22h &    60 &   69 &  77 &  59 & 39 &   46 & 51 & 45 & 62 &   63 &   61 &   63 &  58 \\
  23h &    60 &   68 &  77 &  58 & 39 &   46 & 51 & 45 & 62 &   62 &   61 &   63 &  58 \\
\hline
{Per Month}  &   58  & 68 &  77 &  59  & 40  & 44 &  50 &  45  & 63 & 61 &  62  & 63 & {\bf 58} \\
\hline
\end{tabular}
\end{table*}

\begin{table*}
\caption{Hourly data coverage of radiation from the Texas weather station.
\label{cov-hum-2}}
\begin{tabular}{lrrrrrrrrrrrrr}
\hline
{} & {Jan} &  {Feb} &  {Mar} &  {Apr} &  {May} &  {Jun} &  
                   {Jul} &  {Aug} & {Sep} &  {Oct} &  {Nov} &  {Dec} & {Per Hour} \\
\hline
    6h &    0 &  2 & 29 & 48 & 30 & 45 & 43 & 36 & 37 & 24 & 12 &  0  &   25 \\
    7h &  51 & 80 & 88 & 63 & 31 & 46 & 50 & 59 & 74 & 66 & 67 & 66  &   61 \\
    8h &  60 & 83 & 88 & 64 & 32 & 46 & 50 & 61 & 74 & 66 & 68 & 70  &   63 \\
    9h &  61 & 82 & 88 & 64 & 33 & 46 & 51 & 61 & 75 & 66 & 69 & 70  & 64 \\
  10h &  61 & 82 & 88 & 64 & 32 & 47 & 51 & 61 & 75 & 66 & 70 & 70  &   64 \\
  11h &  61 & 83 & 87 & 64 & 32 & 48 & 52 & 60 & 74 & 67 & 70 & 70  &   64 \\
  12h &  62 & 84 & 87 & 65 & 32 & 48 & 52 & 61 & 73 & 70 & 71 & 70  &   64 \\
  13h &  62 & 85 & 87 & 65 & 32 & 48 & 52 & 63 & 73 & 72 & 72 & 70  &   65 \\
  14h &  63 & 86 & 87 & 65 & 33 & 49 & 52 & 63 & 74 & 72 & 72 & 72  &   65 \\ 
  15h &  63 & 86 & 87 & 66 & 33 & 50 & 52 & 62 & 74 & 71 & 72 & 72  &   66 \\
  16h &  64 & 86 & 88 & 66 & 33 & 50 & 52 & 63 & 73 & 72 & 73 & 72  &   66 \\
  17h &  64 & 86 & 88 & 66 & 33 & 50 & 52 & 63 & 74 & 72 & 67 & 64  &   65 \\
  18h &   7 & 28 & 41 & 39 & 26 & 49 & 52 & 59 & 45 & 12 &  0 &  0  &   29 \\
\hline
{Per Month}  &  62  & 84 &  87 &  64  & 32  & 48 &  51 &  61  & 74 & 69 &  71  & 70 & {\bf 64} \\
\hline
\end{tabular}
\end{table*}

\section{Data coverage}
The data presented here consist of temperature,  atmospheric 
pressure, relative humidity, wind velocities and radiation records acquired 
with the Davis and 
Texas stations using sampling times ranging between 0.5 and 30 minutes. The 
majority of the data were taken with 1 or 5 minutes sampling. Data on wind 
direction and dew points  were also acquired and will  be presented elsewhere, 
in specific studies of the wind characteristics and  atmospheric water vapor
content of the site.

Table \ref{cov_davis_humedad} summarises the  temporal coverage of the 
data, expressed in percentage. The data from the Davis 
station span from October 30, 2000 to February 
18, 2008, with a 70\% effective coverage of the 7.3 year sample: data exists for 
1986 out of 2668 days. The complete sample contains 2693146 minutes;
coverage for day and night are almost equal 1120921  compared to
1122161 minutes. Coverage was $\sim 80\%$ between 2002 and 2006, declining 
to 36\% in 2007. The comparison between the general and the wind data shows that 
bad weather affects our anenometer less than 5\% of the time; in fact our 
logfile indicates that most of our data losses are not due to bad weather but 
to logistics.  

The Texas station started operation a year and a half later and has a
comparable coverage, as shown in Table~\ref{cov_texas_humedad}, with a  58\% 
effective coverage corresponding to  1844977 minutes. 
The data have 3-month gaps in  early 2002 and mid-2003. 
In total we have data for 1584 days out of the 2163  in the period between 
April 12, 2002 and March 13, 2008. The data of the Texas  station have even
coverage for  day and night: 774698  and  763496 minutes,  but are
biased towards the dry period. In Table
\ref{cov_texas_radiation} we present the temporal  coverage of radiation as 
it is higher than for the other parameters, in particular for 2004 the
coverage is almost twice. The complete solar radiation  sample contains 990770  minutes 
from which 526792 minutes are  for the dry season and 463978 for the wet
season.

From tables~\ref{cov-davis-2} and~\ref{cov-hum-1}, it is clear that the coverage per  hour is fairly homogeneous, varying at 
most from 56\% to 59\% for the Texas station, while the coverage per month is 
more variable, specially for the Texas. The coverage per  hour for solar radiation, 
is shown in  table  ~\ref{cov-hum-2}. For the early 
($\sim$6h) and late afternoon ($\sim$18h) hours the coverage is low mostly due to  the
variation of the length of the day, for the other hours the maximum difference is 4\%.

\begin{figure*}
\includegraphics[width=\columnwidth]{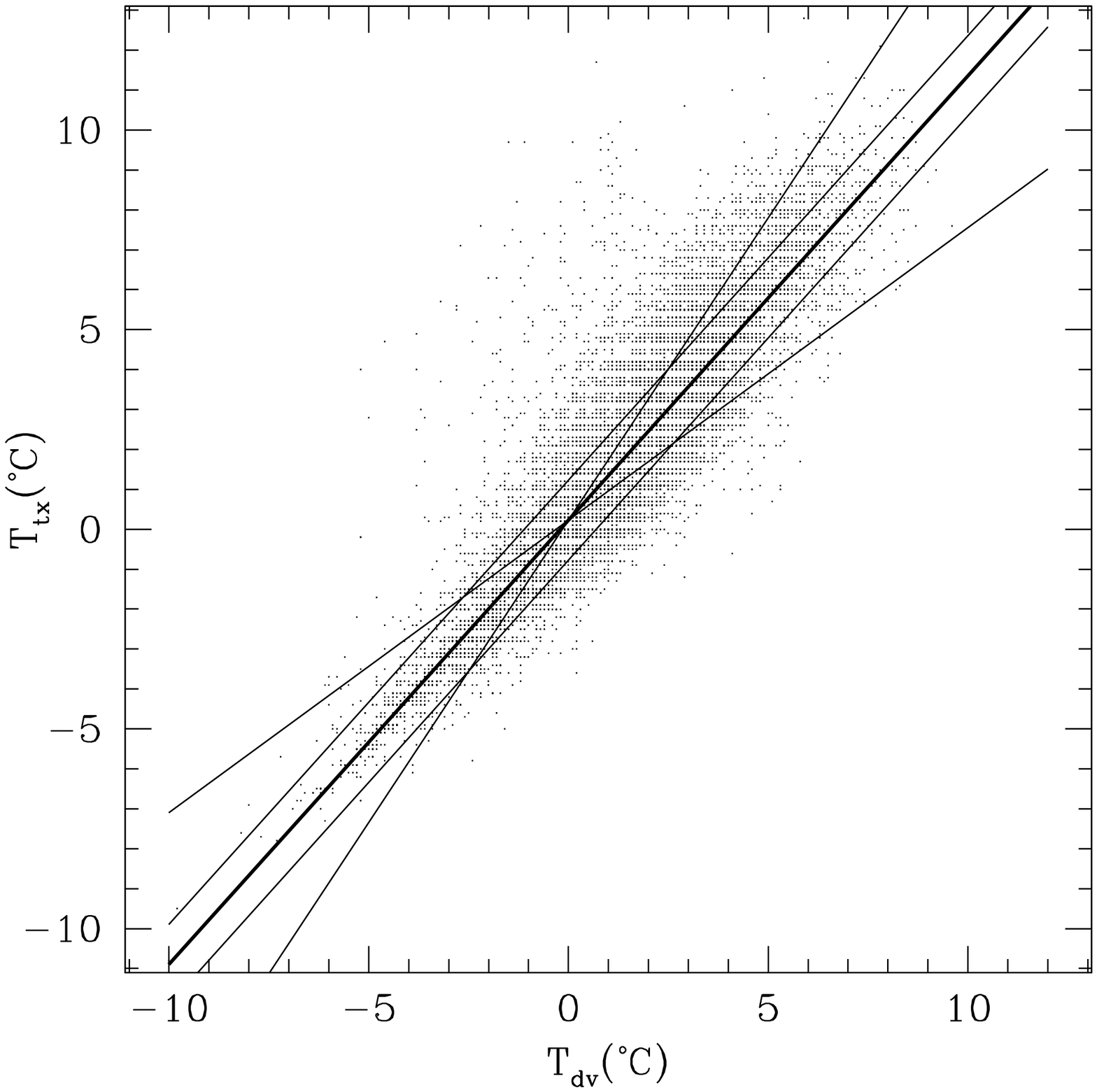}
\includegraphics[width=\columnwidth]{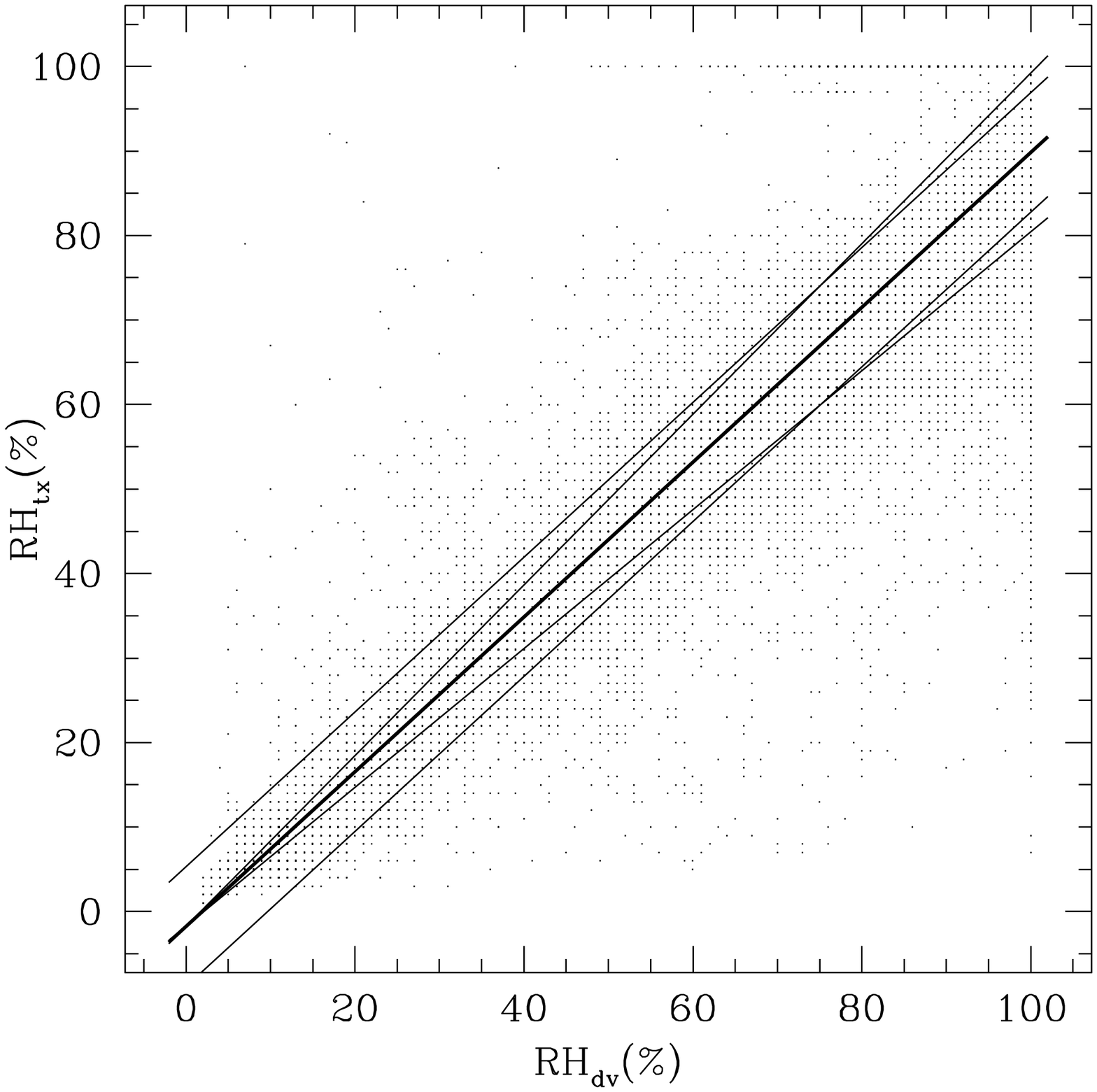}
\caption{Fits between the  Davis and the Texas stations for temperature (right) and RH (left). The fit with the best slope and zero point is indicated with a bold line. The fits adding $\pm 1\sigma$ error to each fitting parameter are also shown: the two parallel fits to the best one correspond to 
$1\sigma$  and $-1\sigma$ in the ordinate; the other two fits correspond to  $1\sigma$ and $-1\sigma$ error in the slope. \label{temp_hum_1a10}} 
\end{figure*}

\section{Cross calibration}

 To  cross calibrate  the two data sets  considering the measurement accuracies 
 we compute  the best linear regression between the two data sets by  minimizing $\chi^{2}$. 
The goodness of the fit is given by  the correlation coefficient  $r$ between the best fit 
and the data points. 

  The plots with all  the data points  and the best fit have more than $10^{5}$ points for 
each parameter ($>$2MB). 
To display them can be misleading because they tend to look as scatter plots. They would suggest that the fitting
errors are underestimated as it is impossible to distinguish if in a given position there  is one
or $10^{3}$ points. We decided to present plots with 1/10 of the data points randomly  chosen as in this case the 
points tend to be located where the density of data is larger and, on the other hand,  the files are handled. 

For a given plot, the equality  between the Davis  and  Texas data  corresponds to a straight line 
at 45$^{\circ}$, not shown. We present the best fit and zero point (bold line) and the fits obtained adding the $1\sigma$ error  to each parameter.  

We report the fits  -and the statistics-  with two decimal digits as the determination of a statistical value can
be made with higher accuracy than the nominal resolution of the instrument  for data with high signal to noise 
ratio (S/N).  In our analysis  the high S/N  is due to a very large  data set.

 Subscripts $ _{dv}$ and $ _{tx}$ stand for the Davis and Texas stations, respectively, for any of the compared parameters: temperature $T$ in $^\circ C$, relative humidity $\mathrm{RH}$ in per cent, and wind speed $w$ in m~s$^{-1}$.

\subsection{Temperature}
We compared temperature data from both stations registered with same times, allowing 
for up to a one minute difference between clocks. The 233985 registers common to
the Davis and Texas station are in fair agreement, with a correlation of 0.90 for
the linear fit,  
\begin{equation}
T_{tx} = (1.11\pm 0.4) \,T_{dv} + (0.23\pm 1.0)^{\circ}\rm C , \label{t-fit}
\end{equation}
 the rms scatter around the fit is  $1.23^{\circ}\rm C$, which is 1.7$\sigma$, where 
$\sigma$=0.7 is the combined error accuracy  of both stations. The data points with the best fit
and the fits obtained adding $\pm 1\sigma$ are shown on the left side of Fig. \ref{temp_hum_1a10}.

Relation~(\ref{t-fit}) leads to a larger range of temperatures for the Texas station 
than for the Davis one and 
differences $\sim 1^{\circ}\rm C$ for {\em extreme} temperature,
$|T|\ga 10^{\circ}\rm C$. 
Additionally, the comparison between the Davis station and the Backup
station gives a fit with a slope of 1.03, intercept of 0.03 and  $r=$0.99.
We consider the temperature data from different stations to be  consistent with each other.

\subsection{Relative humidity}
Even though both stations have similar humidity sensors, we found significant 
systematic differences in simultaneous measurements. Some of these might be 
attributed to local differences in humidity, due to fog moving accross the site,  
as the stations are located  130~m apart. The common data are well correlated 
($r=0.94$) with a linear fit marginally consistent with a one to one relation, 
\begin{equation}
{\rm RH}_{tx} = (0.92\pm 0.09)\,{\rm RH}_{dv}-(1.8\pm 7.1)\% . \label{h-fit}
\end{equation} 
The values from the Davis station tend to be 10\% higher than those from the Texas. 
The comparison between the Backup and Davis stations gives a similar 
fit, of slope 0.91, with a correlation equal to 0.99, being more consistent with the 
data from the Texas weather station. The data points with the best fit (bold line)
and the fits obtained adding $\pm 1\sigma$ are shown in Fig. \ref{temp_hum_1a10}, 
on the right hand side.

Of further significance is the variation of this fit with time, shown in 
Table~\ref{hr-slopes}. The fits for 2002 and 2003 are compatible with equal 
measurements from both stations,  ${\rm RH}_{tx} = {\rm RH}_{dv}$, while 
those onwards from 2005 become deviant.
The measurements of the stations were similar at earlier times and diverged with 
time. The slope  drifted from $0.97\pm 0.10$ in 2002 to $0.82\pm 0.10$ 
in 2007, resulting in larger humidity measurements for the Davis station, by about 
10\%.  In fact, the Davis weather station data show such a trend. The best fit of the
yearly median ${\rm RH}_{dv}$ {\em vs.} year shows a 4\% increase per year with 
a correlation coefficient of 0.91.  However,  the Texas
station does not  show this trend  as the best fit of  the yearly
median  ${\rm RH}_{tx}$ {\em vs.}  year  shows a slope
of $-2.1\%$ per year  and a correlation  coefficient of  $-0.25$.

\begin{table}
 \centering
  \caption{Linear fits, RH(Texas) = slope * RH(Davis) + intercept, to common 
data on relative humidity.\label{hr-slopes}}
  \begin{tabular}{lrrrrr}
  \hline
Sample   & slope  & intercept & rms & correl & N$_{\rm points}$\\
         &        &     \%    &  \% &        &  \\  
         &        &     $\pm 7.07$    &    &        &  \\  
  \hline
  2002 & $0.97\pm 0.10$ & $-1.00$  & 11.5 &  0.94  &  72025 \\
  2003 & $0.94\pm 0.10$ & $-2.22$  &  9.0 &  0.96  &  52103 \\
  2004 & $0.90\pm 0.10$ & $-1.90$  &  9.2 &  0.96  &  12705 \\
  2005 & $0.88\pm 0.08$ & $-4.27$  & 11.2 &  0.88  &  33024 \\
  2006 & $0.89\pm 0.09$ & $-4.63$  &  9.1 &  0.96  &  36622 \\
  2007 & $0.82\pm 0.01$ & $-0.82$  & 11.8 &  0.93  &   3284 \\
  2008 & $0.89\pm 0.09$ & $-5.78$  &  8.9 &  0.94  &   2354 \\
\hline
  All  & $0.92\pm 0.09$ &  $-1.77$  & 11.2 &  0.94  & 212117 \\
\hline
\end{tabular}
\end{table}

\begin{figure*}
\includegraphics[width=\columnwidth]{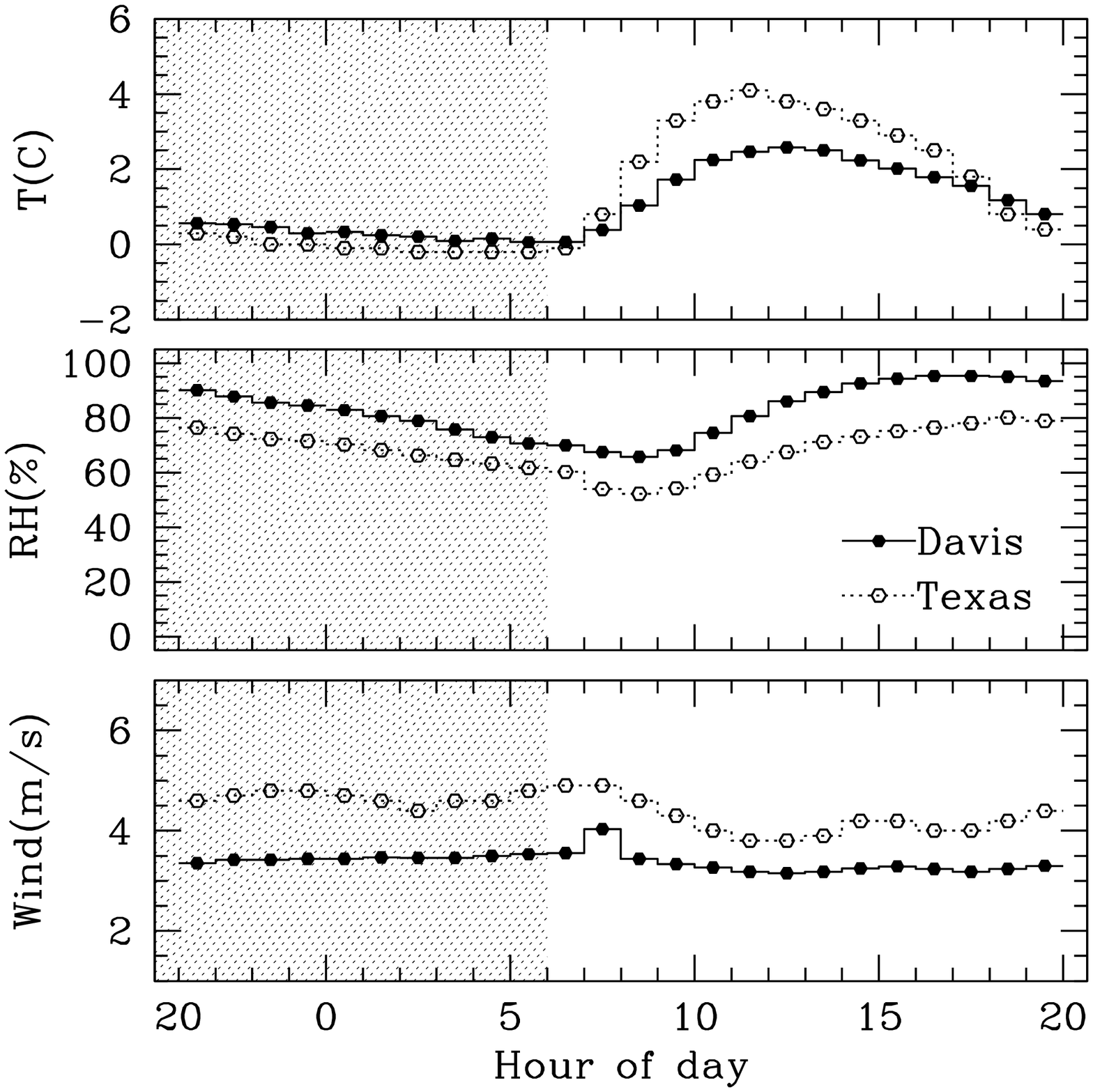}
\includegraphics[width=\columnwidth]{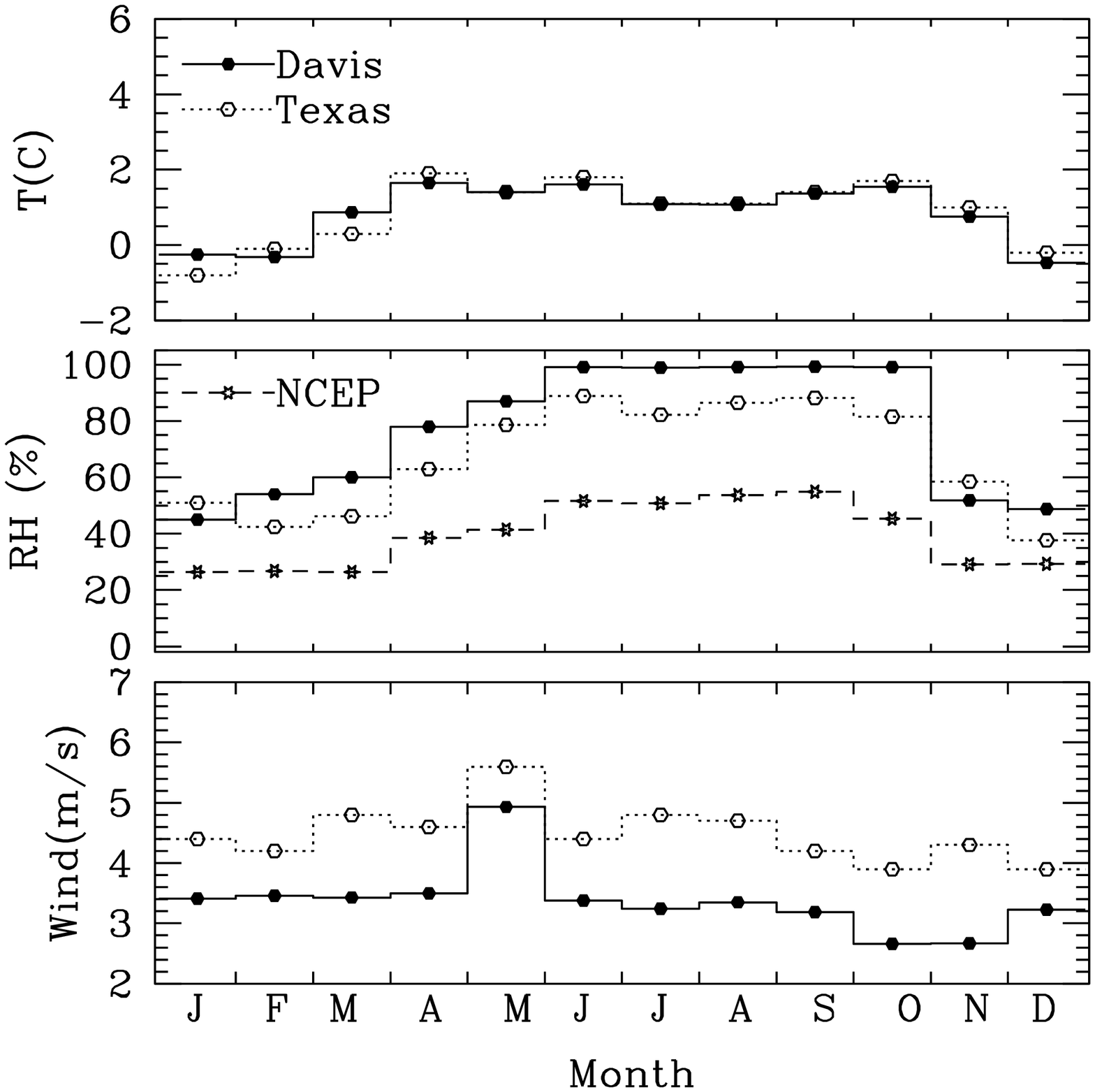}
\caption{Comparison between data from the Davis and Texas weather stations. 
Left: median values per time of day. Solid lines and filled circles represent
the Davis stations; open circles and dotted line Texas. Here, the RH very low  values 
 measured during the dry season,  are blended with the high values during the wet season. 
 Right: the same for monthly values, in the RH plot the curve with stars corresponds 
to the NCEP Reanalysis model  for a 600 mbar pressure level.  }\label{compara}
\end{figure*}

To complement   the relative humidity analysis we used the RH monthly mean data 
at 600 mb provided by the National Center for Environmental Prediction/National 
Center for Atmospheric Research (NCEP/NCAR) Reanalysis project. 
The NCEP data, included in Fig.~\ref{compara} in dotted lines, follow the same 
trend as those measured in situ, but with lower values by up to 40\%. The
 offset can be explained by the fact that some variables as RH are partially 
defined by observations but also are strongly influenced by the local topography
and the characteristics of the NCEP analysis model, as pointed out by~\citet{ncep}.
Even with an offset these data can be used to look for a tendency. 
In the case of annual trend,   the NCEP reanalysis data does point to an increase 
of RH of 0.9\% per year with a correlation coefficient of 0.7. The data actually 
shows a 1\% per year decrease in RH from 2001 (39\%) to 2004 (36\%) followed
by increased mean values of 42\% between 2005 and 2007.  This increase is not
comparable to the Davis one. We conclude that the RH sensor of the Davis station 
drifted with time proving higher values than the real ones.

\begin{figure}
\includegraphics[width=\columnwidth]{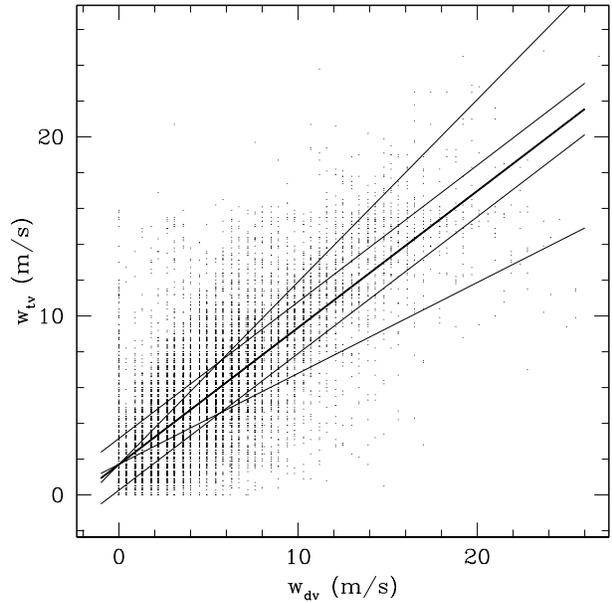}
\caption{Wind speed:  fit with a best slope and zero point is indicated as a bold line. The fits adding   $\pm 1\sigma$ error to each fitting parameter are also shown: the two parallel fits to the best one correspond to $1\sigma$  and $-1\sigma$ in the ordinate; the other two fits correspond to  $1\sigma$ and $-1\sigma$ error in the slope. \label{viento_1a10}}
\end{figure}

\begin{figure*}
\includegraphics[width=\columnwidth]{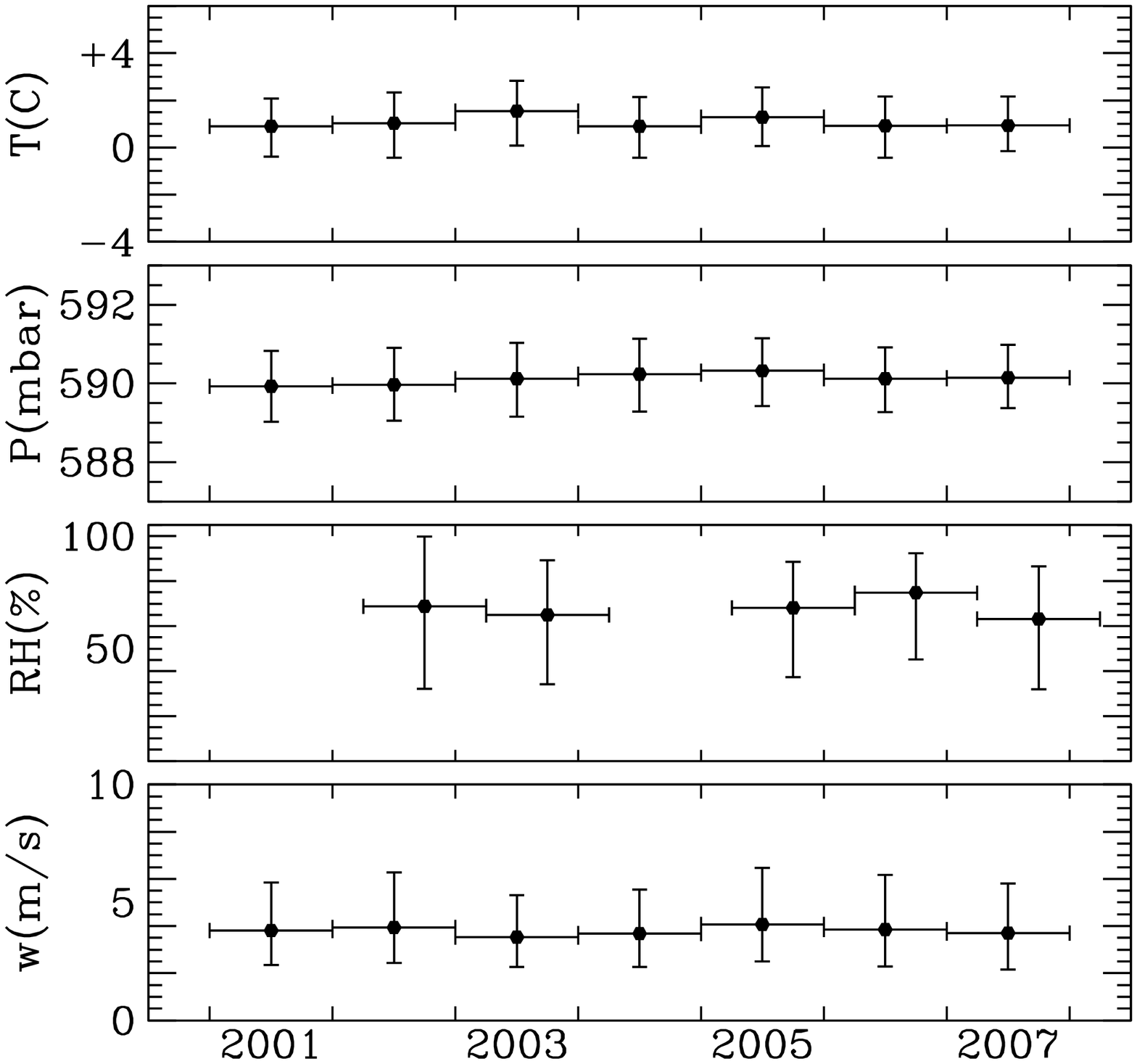}
\includegraphics[width=\columnwidth]{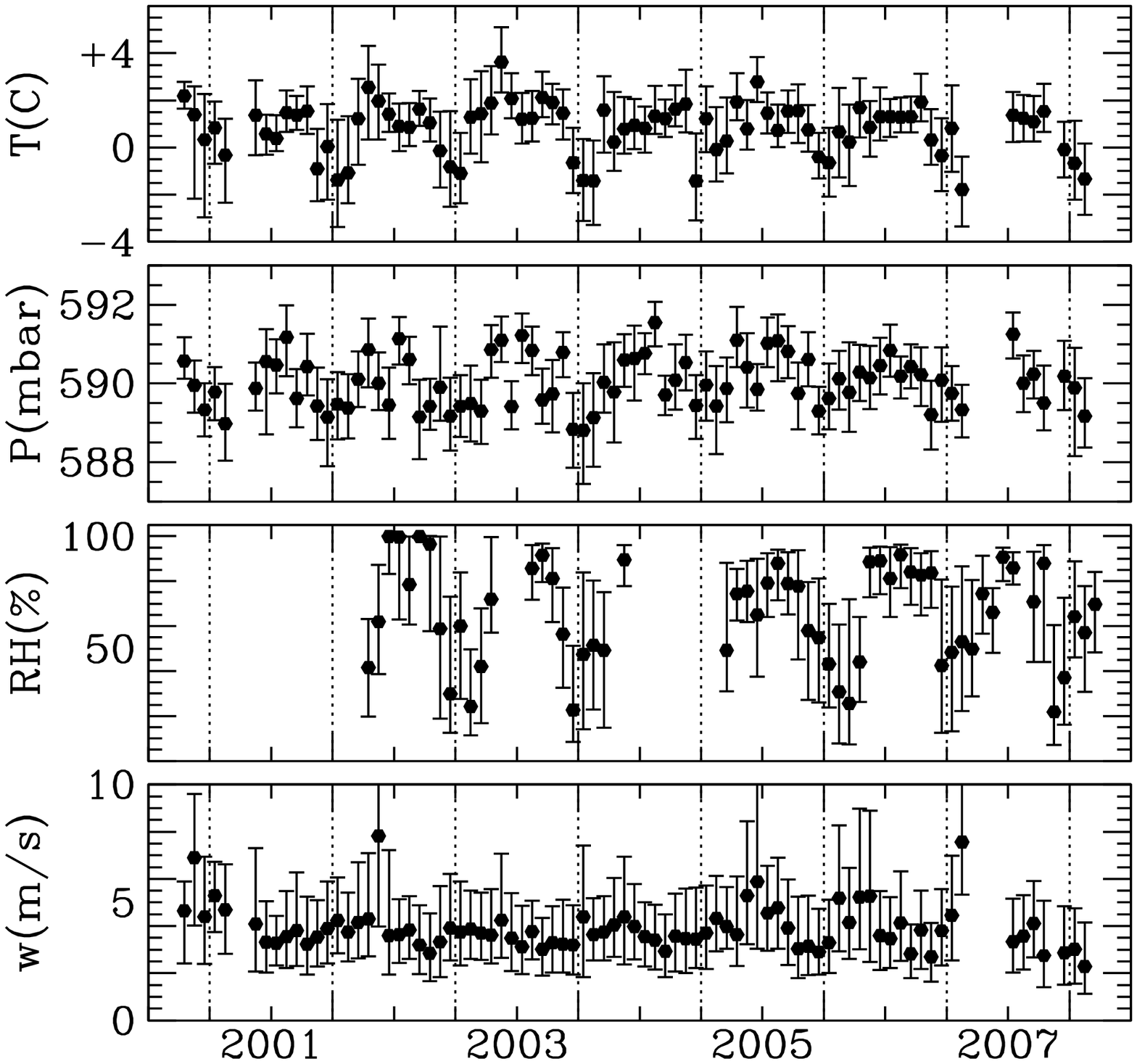}
\caption{Statistics for all the data,  points are medians with 
bars going from the first to the third quartile. Left: yearly statistics.
Right:  monthly statistics. The stability of the parameters is appreciated
in the annual median values. However more information is extracted  from the
monthly statistics, in particular in the case of RH that is very low during
the dry seasons months as  we discussed in  detail in the  RH section.
\label{anuales_monthly}} 
\end{figure*}

\begin{figure*}
\includegraphics[width=\columnwidth]{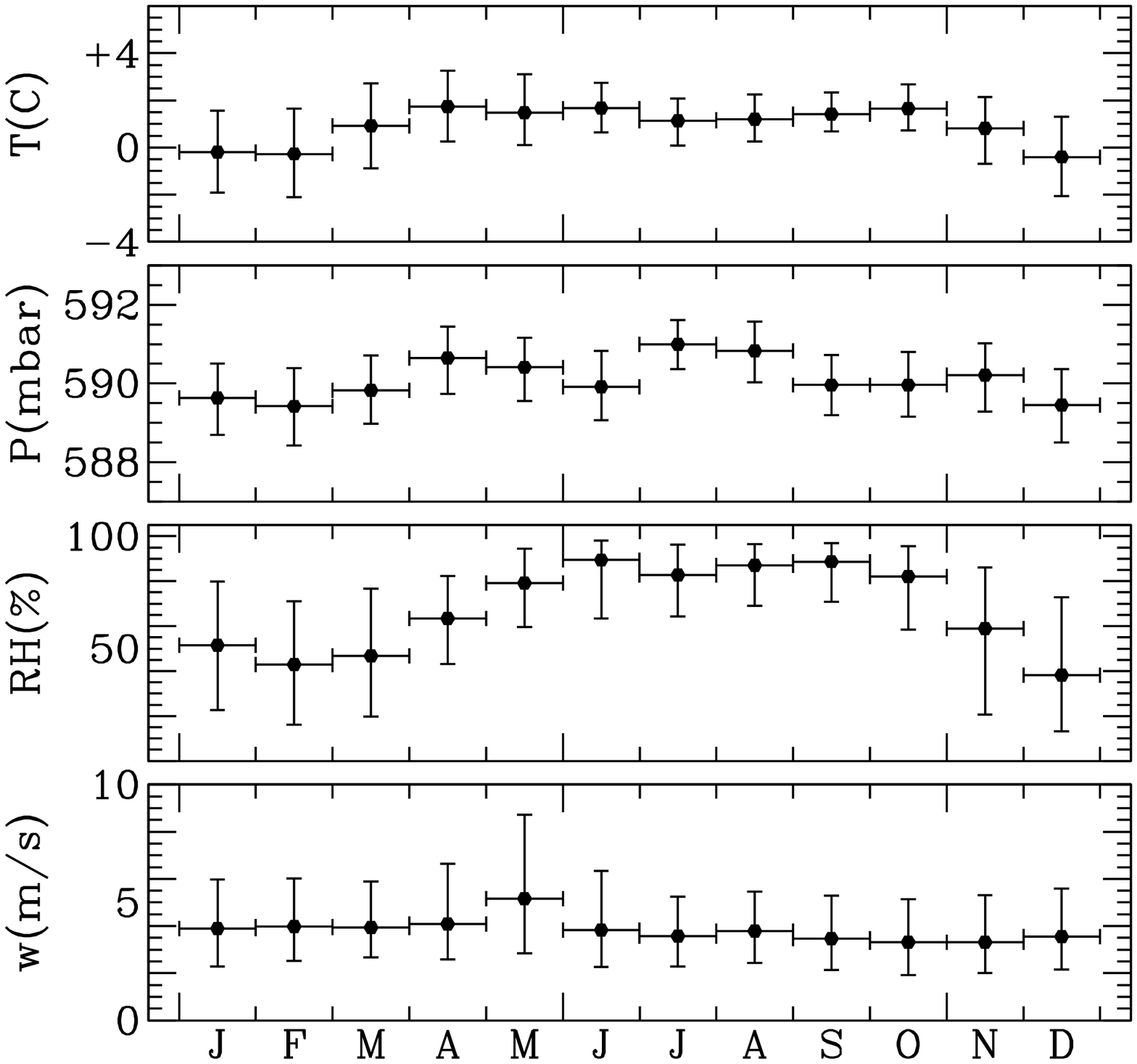}
\includegraphics[width=\columnwidth]{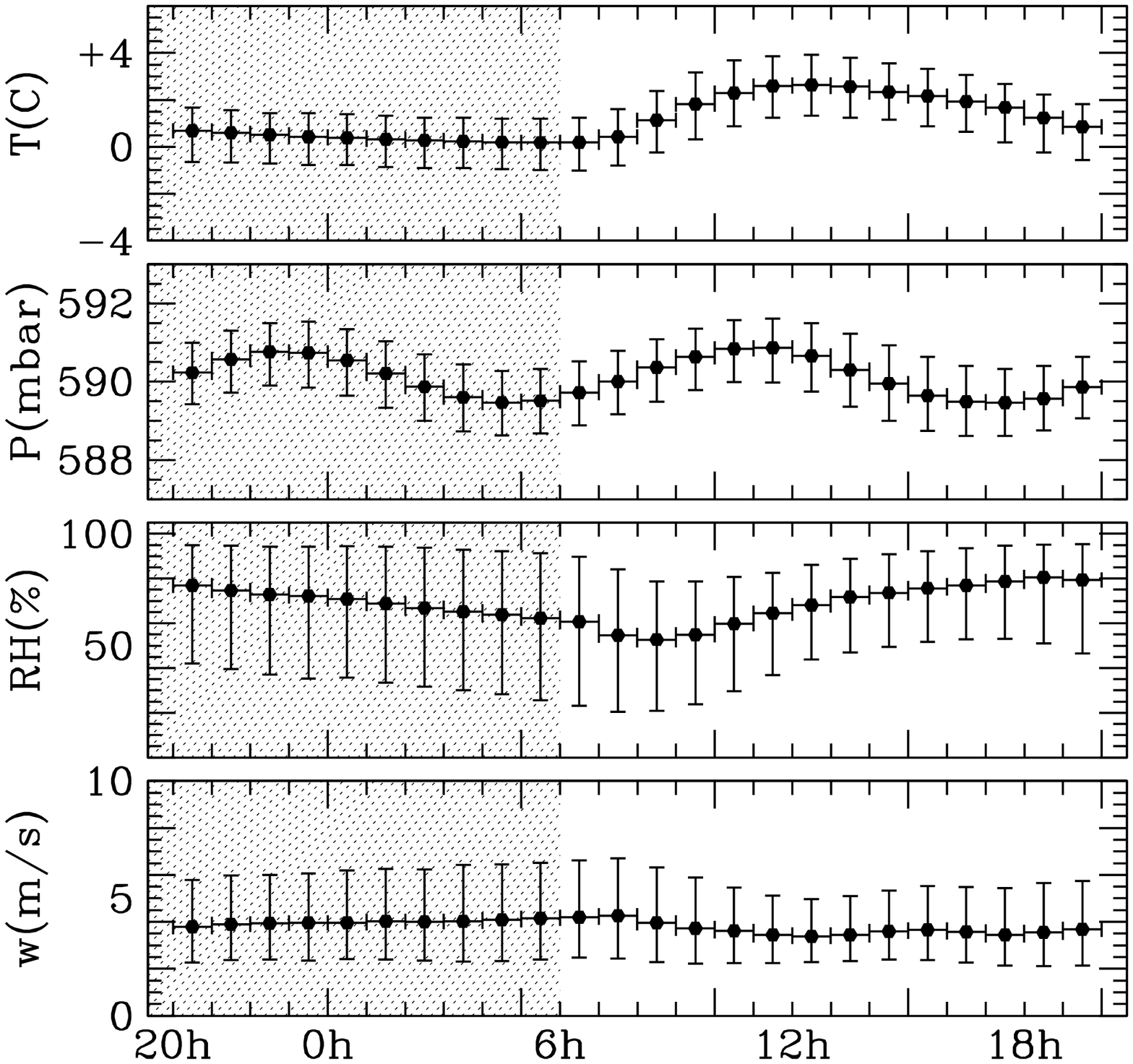}
\caption{ Left: monthly statistic. Each point corresponds to the  monthly median 
for all the  years  with the  bars going  from the 
first to the third quartile. 
RH is the parameter that shows the strongest seasonal
dependence.  Right: hourly statistics of all data. The points are the
median values  and the  bars go from the first to the third quartile. The
daily cycle is recognizable for all the parameter but in the case of RH
the very low  values  measured during the dry season,  are 
blended with the high values during the wet season. } 
\label{monthly_hour}
\end{figure*}

\subsection{Wind velocity}
Simultaneous wind measurements from both stations follow the
best fit,
\begin{equation}
w_{tx}= (0.76\pm 0.26)\,w_{dv}+(1.7\pm 1.4){\rm m~s^{-1}} ,
\label{wind-fit}
\end{equation} 
with a correlation coefficient of 0.75. The simultaneous outputs from the 
anemometres are less correlated than the temperature and humidity data. 
The Davis weather station tends to give lower values. The fit marginally
excludes $w_{tx}=w_{dv}$ in the parameter  1$\sigma$ error contour. 
The rms deviation of the fit is 2.29~m~s$^{-1}$, more than 
twice the combined measurement error of 1.02~m~s$^{-1}$. The data points 
with the best fit and the fits obtained adding $\pm 1\sigma$ are shown in
 Fig. \ref{viento_1a10}.

The median velocities of both stations are below 
the LMT operation  threshold of 9$~m~s^{-1}$. Furthermore, looking at the  
distributions  we note that the percentage of wind speeds below 
the LMT operation threshold are very similar for both stations. 
We compared the best fit with the fit $w_{tx}=w_{dv}$ and we found
that the zero point in the best fit cancels the effect of the 
different slope giving the same  statistical behaviour of both data, 
although not necessarily simultaneous  values. The fact that the
statistical behaviour of the two data sets is similar  supports the
premise  that there can be genuine differences in wind speeds
due to the topography of the  mountain top.

\section{Data analysis and results}

Median and quartile values were computed for the different parameters and 
datasets. The data points were 
weighted by sampling time.  A time interval larger than 16 minutes 
without data  is considered a gap. We did not compensate for gaps in the data.

The samples were divided in different subsamples for their study, 
defined as follows:  {\em daytime} is the 10 hr interval between 8:00am and 5:59pm; {\em nighttime} 
ranges from 8:00pm to 5:59am. These two definitions avoid  the transition times 
of sunrise and sunset to analyse the weather under stable conditions. 
The {\em dry season} is the 181 day period from November 1st to April 30; the 
{\em wet season} goes from May 1st to October 31st, covering 184 days. The time 
span period of the Davis data comprises 1378 dry season days (978 with data) 
and 1290 wet season days (1008 with data); the time span period of the Texas 
data comprises 1059 dry season days (889 with data) 
and 1104 wet season days (695 with data).  


\begin{table}
 \centering
  \caption{Weather data statistics. Temperature, pressure and wind speed are
from the Davis station while relative humidity is from the Texas weather station,
as explained in the text. Quartiles are estimated with an accuracy of $\Delta x/100$,
where $\Delta x$ is the sampling of the parameter under consideration. \label{global}}
\begin{tabular}{lcrrrrrrr}
\hline
   & cov(\%)& min &  $q_1$ & median & $q_3$  & max\\
\hline
\multicolumn{4}{l}{\underline{ Temperature ($^\circ$C)}}\\
All    &  71  & --10.6 & --0.30 &  1.07 &   2.32 &  11.8 \\
Dry    & 71   & --10.6 & --1.42 &  0.34 &  2.10 &   9.9 \\
Wet    & 72   &  --4.8 &  0.38 &  1.39 &  2.48 &  11.8 \\
Day    &  71  & --10.4 &  0.73 &  2.12 &  3.36 &  11.8 \\
Night  & 71   & --10.2  & --0.82 &  0.35 &  1.37  &  6.7 \\
\hline
\multicolumn{4}{l}{\underline{Atmospheric pressure (mbar)}}\\
All        &  71 & 580.2  & 589.19 & 590.11 & 590.99 & 594.7 \\
Dry      &  71 & 581.6  & 588.85 & 589.82 & 590.74 & 594.3\\
Wet      &  72 & 580.2  & 589.51 & 590.39 & 591.19 & 594.7 \\
Day      &  71 & 580.8  & 589.26 & 590.22 & 591.14 & 594.7 \\
Night    &  71 & 580.2  & 589.22 & 590.13 & 591.00 & 594.7 \\
\hline
 \multicolumn{4}{l}{\underline{Relative humidity (\%)}} \\
All         & 58 & 1 & 36.73 & 68.87 & 92.59 & 100 \\
Dry       & 65 & 1 & 20.82 & 50.92 & 78.52 & 100 \\
Wet       & 51 & 2 & 64.86 & 84.92 & 96.18 & 100 \\
Day       & 58 & 1 & 40.70 & 68.19 & 88.73 & 100 \\
Night     & 58 & 1 & 33.18 & 69.35 & 93.96 & 100 \\
\hline
 \multicolumn{4}{l}{\underline{Wind speed (m s$^{-1}$)}}\\ 
All   & 69 &  0 &  2.31 &  3.77 &  5.88 & 36.2 \\
Dry   & 68 &  0 &  2.36 &  3.80 &  5.91 & 36.2 \\
Wet   & 70 &  0 &  2.27 &  3.74 &  5.85 & 35.8 \\
Day   & 69 &  0 &  2.28 &  3.57 &  5.45 & 35.8 \\
Night & 68 &  0 &  2.36 &  3.98 &  6.18 & 35.8 \\
\hline
\end{tabular}
\end{table}


\subsection{Parameter statistics - seasonal and diurnal}
  
  The data statistics 
of the meteorological parameters, except solar radiation, 
are summarised in table~\ref{global}, for the entire samples and for the 
day/night, dry/wet season subsamples. The columns indicate from left to right,
coverage  percentage,  minimum,  first, second, third quartile values
($q_1, q_2, q_3$) and maximum. Temperature,  atmospheric pressure and 
wind speed statistics are from the Davis data,  while statistics for 
relative humidity statistics were computed for the Texas weather station. 
The ``all'', ``day'' and ``night'' statistics were estimated compensating for the 
uneven seasonal coverage of the Texas data, using 695 days out of the 889 days 
dry season data.  
 For solar radiation, the subdivision ``day'' and ``night''   become meaningless.  
The results of the analysis 
of solar radiation data, including the influence of daytime cloud cover, are 
described in subsection~\S\ref{radiacion}.

Table~\ref{annual} displays the annual statistics of each parameter, considering 
also the dry and wet seasons. In Fig.  \ref{anuales_monthly} we plot the median 
values of the measured parameters per year and per month, with  bars going 
from the first to the third quartile. Fig.~\ref{monthly_hour} shows the statistics 
of the data folded per month and per hour of day, in order to show the seasonal 
and daily modulations.

\begin{table*}
\centering
\caption{Median values of the meteorological parameters per year and  season 
\label{annual}.}
\begin{tabular}{lrrrrrrr}
\hline
Parameter                  & 2001      & 2002  & 2003 & 2004 & 2005 & 2006  & 2007 \\
\hline
Temperature  ($^\circ$C)   &   0.90    &  1.02 & 1.54 & 0.89  & 1.28 & 0.91 & 0.93 \\
Dry season                 & $-0.10$   &  0.25 & 0.80 & 0.10  & 0.87 & 0.32 & 0.08 \\
Wet season                 &   1.25    &  1.28 & 1.93 &  1.18 & 1.57 & 1.35 & 1.25 \\
\hline
Pressure (mbar)            & 589.93 & 589.97 & 590.12 & 590.24 & 590.32 & 590.12 & 590.15 \\
Dry season                 & 589.38 & 589.81 & 589.85 & 589.75 & 590.12 & 589.89 & 589.84 \\
Wet season                 & 590.29 & 590.11 & 590.42 & 590.59 & 590.47 & 590.40 & 590.40 \\
\hline
Relative Humidity (\%)    &  & 68.86 & 65.06 & (74.50) & 68.10 & 74.80 & 63.13 \\
Dry season                &  & 42.01 & 53.53 & (49.21) & 49.94 & 52.69 & 58.30 \\
Wet season                &  & 99.56 & 86.07 & (89.50) & 78.95 & 85.28 & 81.47 \\
\hline
Wind (m~s$^{-1}$)         & 3.81 & 3.94  & 3.54  & 3.68  & 4.07  & 3.86 & 3.71 \\ 
Dry season                & 4.28 & 4.01  & 3.60  & 3.78  & 3.74  & 3.95 & 3.81  \\
Wet season                & 3.49 & 3.88  & 3.48  & 3.60  & 4.42  & 3.74 & 3.66  \\
\hline
\end{tabular}
\end{table*}


\begin{figure*}
\includegraphics[width=\columnwidth]{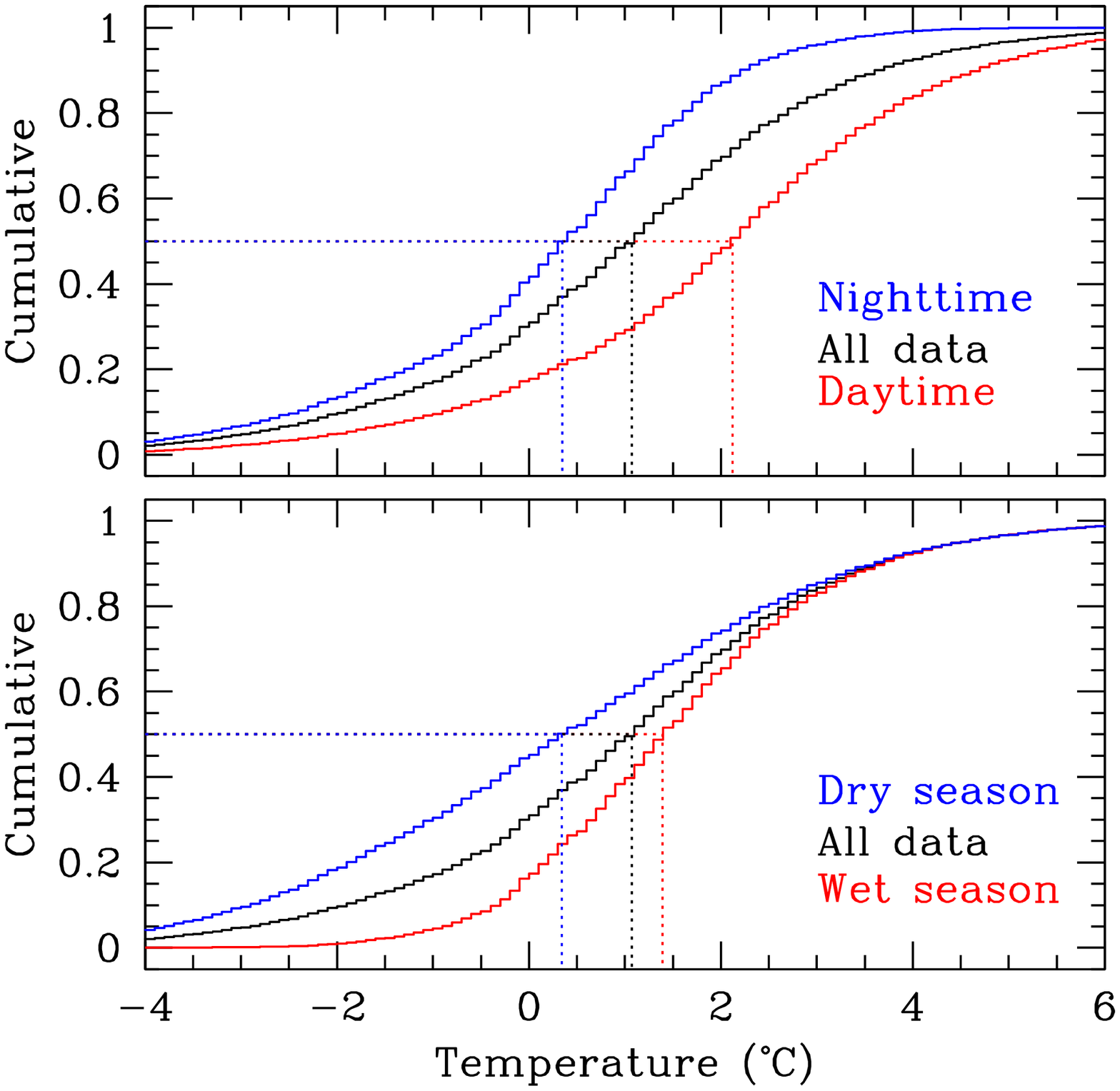}
\includegraphics[width=\columnwidth]{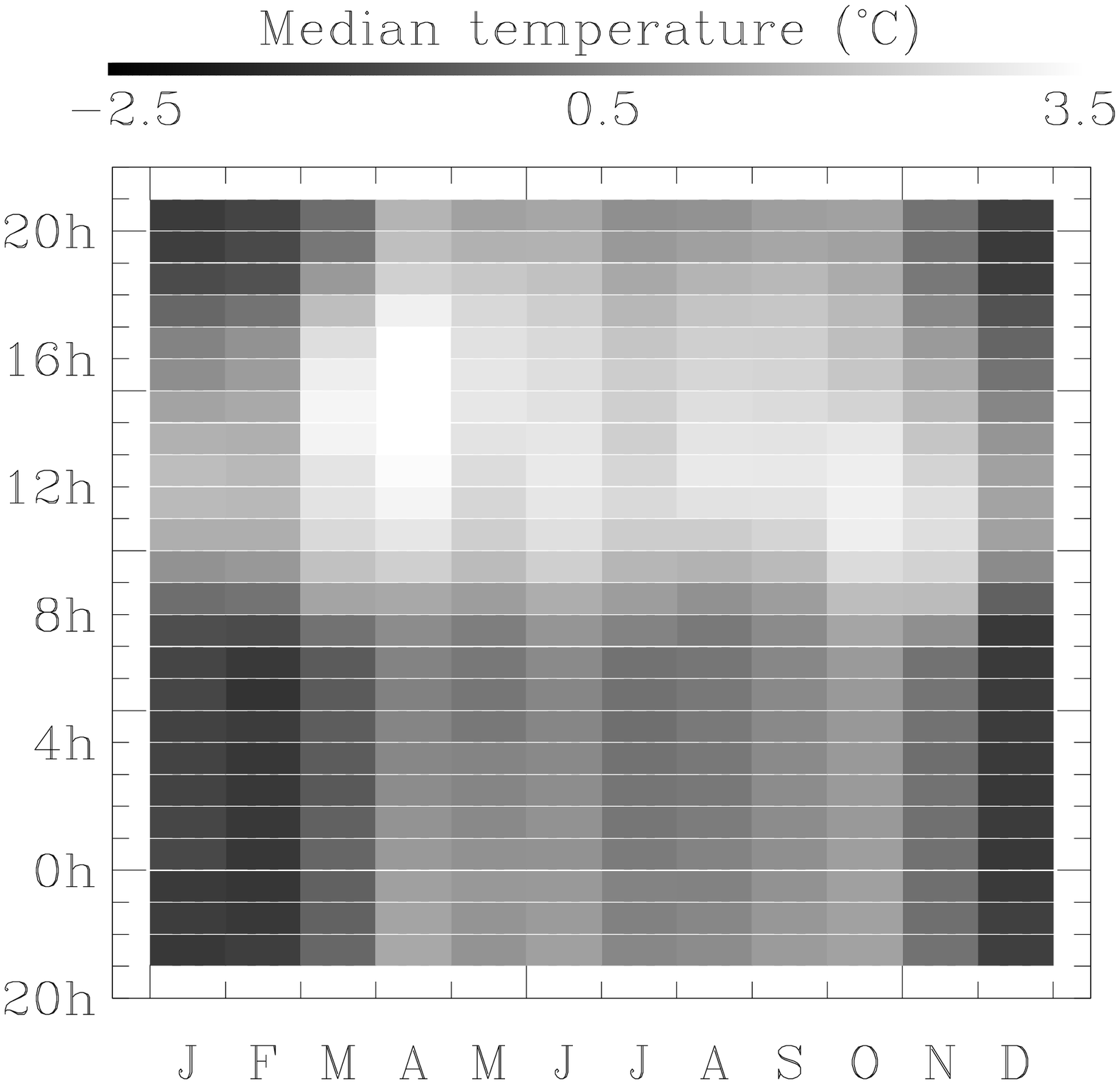}
\caption{Left: the temperature distribution for the whole data set is shown in black with a median
of $1.07^{\circ}$C. 
The distributions for nighttime, daytime, dry and wet season are shown as indicated. The median
values are  $0.35^{\circ}$C, $2.12^{\circ}$C, $0.34^{\circ}$C and $1.39^{\circ}$C respectively.  Right: 
a three dimensional representation of the  temperature behaviour where the grey 
scale intensity corresponds to the temperature median  for a given month and hour.  
The annual and daily cycle are apparent in the horizontal and vertical structure of 
the figures, behaving as expected. See the electronic edition of MNRAS 
for a color version of this figure.}\label{temp3dq2} 
\end{figure*}

\subsection{Temperature}
According to the Davis weather station, the median temperature  for the site is 
$ 1.07^{\circ}$C, with quartile values of $-0.30^{\circ}$C and $ 2.32^{\circ}$C. 
 The extreme temperatures recorded by the Davis station 
on site are  relatively  mild: the minimum temperature in the data 
is $-10.6^{\circ}$C while the maximum   11.8$^{\circ}$C. 
The Texas station registered the same median temperature, but 
with somewhat larger variations and more marked extremes:
$-13.3^{\circ}$C and  $ 14.4^{\circ}$C.   As it would be expected that
both stations register the same extreme temperature, it is required to perform
some experiments to determine if these differences are real or are due
to the distinct temperature sensors sensitivity to extreme conditions.  We plan
to carry out  such experiments.  In any case the 
temperatures at the site do not show large variations.

The daily cycle,  quantified as the difference between the night 
and day medians, is 1.77$^{\circ}$C, going from 0.26$^{\circ}$C to 2.03$^{\circ}$C respectively.  A similar value is obtained for seasonal variations: the median and third quartile 
($q_3$) values for dry/wet differ by only 1.76$^{\circ}$C and 1.09$^{\circ}$C respectively. The lowest quartile does show a larger -but still mild- difference, 
close to  2$^{\circ}$C. In Fig.~\ref{monthly_hour}  the amplitude of the curve  between
the lowest median, -0.22$^{\circ}$C at 5am, and the highest median, 4.15$^{\circ}$C 
at 11pm,  is 4.37$^{\circ}$C. The coldest month is  December, with a median
 of $-0.59 ^{\circ}$C,  2.5$^{\circ}$C below the warmest month, June, with a median of 
$ 1.91^{\circ}$C.  

Temperature distributions are shown in the cumulative histograms on the left 
panel of Fig.~\ref{temp3dq2}. The distributions for  nighttime, daytime, dry 
and wet seasons are shown as indicated. The temperature differences due to the
diurnal cycle are larger for values above the median while the seasonal temperature
differences are larger  for values below the median.  The right hand side 
panel of the same figure shows in a grey scale diagram the medians per hour 
and month. Temperatures are at their lowest during the nights of the dry months, 
specifically between December and February, and highest around or just after 
noon between April and June. Note that the period between July and September 
is not warmer than April and May, due to the effect of rain.
 
If we consider the altitude of the site and the temperature gradient of a standard atmosphere 
model, $dT/dz=-6.5^{\circ}\rm C/km$, the corresponding sea-level temperature
is $T_{0}=30.9^{\circ}$C, about  $16^{\circ}$C above the standard atmosphere
base value. This is clearly an effect due to the low latitude, which results 
in a warmer temperature at a high altitude site.
A final remark is that the site presents a  good degree of thermal stability,
beneficial for scientific  instruments: thermal stability will help the performance of the LMT, 
designed to actively  correct its surface to compensate for gravitational and thermal deformations.

\begin{figure*}
\includegraphics[width=\columnwidth]{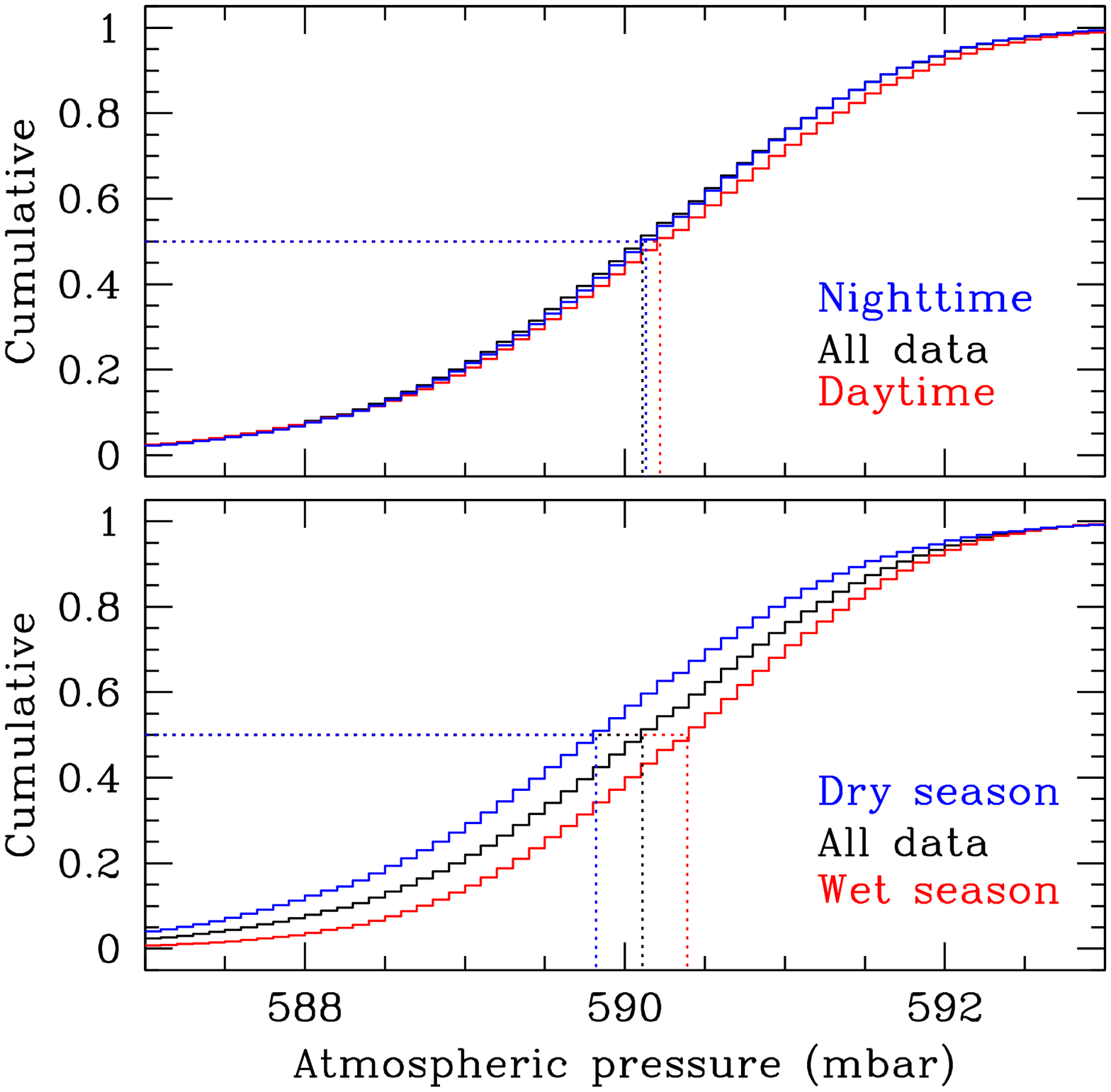}
\includegraphics[width=\columnwidth]{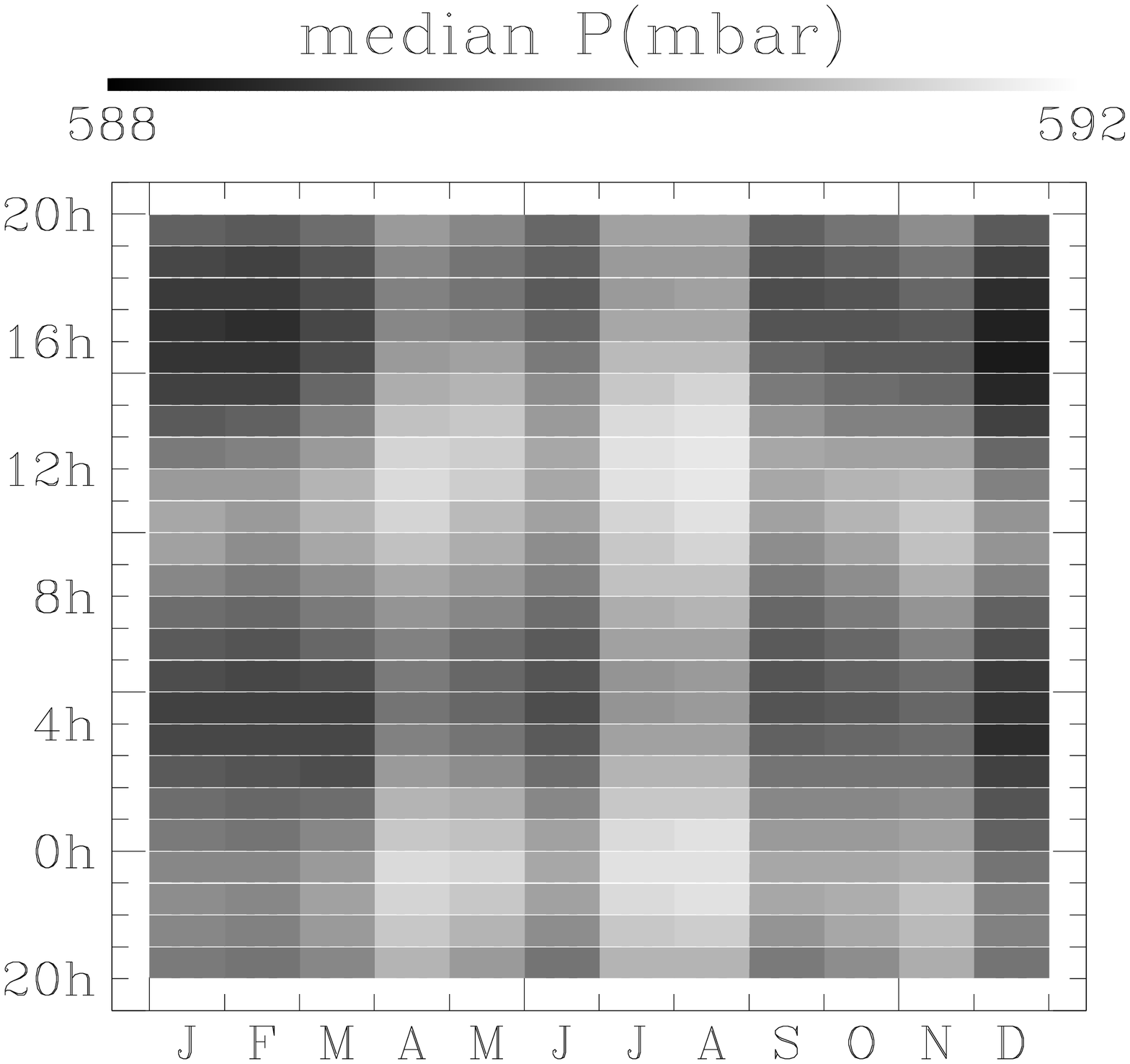}
\caption{Right: The atmospheric pressure distribution for the whole data set is shown 
in black with a median  of 590.11 mbar.  The distributions for nighttime, daytime, dry and 
wet season are shown as indicated. The median values are: at nighttime  590.13 mbar, at 
daytime 590.22 mbar, during the dry season 589.82 mbar and during the wet season  590.39 mbar.  
Left: a three dimensional representation of the atmospheric pressure, the intensity corresponds to
the atmospheric pressure median value for a given month and hour. The daily and yearly cycle are 
clearly recognizable. See the electronic edition of MNRAS 
for a color version of this figure.
\label{pres3dq2}} 
\end{figure*}

\subsection{Atmospheric pressure}
The barometre of the Texas weather station did not provide meaningfull data
and these had to be discarded. We discuss only the data from the Davis weather
station. To verify the calibration of the Davis barometre we performed a comparison
with a basic water barometre,  for about 3 hours during daytime, obtaining  a pressure 
of 594~mbar, in very good  agreement with the  Davis barometre reading of 592.4~mbar.
Therefore,  the Davis weather station gives readings accurate 
to within 2.4~mbar.

The site presents a low atmospheric pressure which is  characteristic of a high 
altitude site. The median is 590.11~mbar with a daily cycle of 1.45~mbar, as 
measured by the difference between the median of the 4~am sample (589.36~mbar) 
and that of the 11am data (590.81~mbar), displayed in Fig.~\ref{monthly_hour}. 
The daily cycle is in fact a double 12 hour cycle, with maxima at 11h and 23h
and minima at 5h and 17h.   This semidiurnal pressure variation of a
few mbar  is well  known for low latitude zones.  It  is associated with 
atmospheric tides excited by heating due to  insolation absorption by 
ozone and water vapor \citep{Lindzen79}.  The yearly
cycle is not as  well defined, see Fig.~\ref{annual},  with relative
minima in February (589.37~mbar), June and December, and maximum value in July
(590.87~mbar), for a peak to peak amplitude of 1.5~mbar. 

High and low pressure are usually related to good and bad weather, respectively.
The largest pressure recorded on site is 597.4~mbar, 3.8 ~mbar above the median,
just before midnight on the 17/8/2001 and again at noon 18/8/2001. The weather 
was dry  as relative humidity values were  18\% and 22\% respectively with 
temperatures 
of 1.4$^{\circ}$C and 4.7$^{\circ}$C. We note that while the weather is usually 
poorer in the wet season, these good conditions happened in August, indicating 
that good observing conditions can happen any time of year; the largest atmospheric 
pressure during the dry season occurred on 7/3/2004 at  10:35~am, when the weather 
record indicated a pressure of 594.3~mbar, temperature of 5.9$^{\circ}$C
and a relative humidity of just 12\%. 

The lowest pressure corresponded to what has presumably be the worst weather on 
site: the relatively close passage of hurricane Dean, on 22/8/2007. At 4:30am 
when the pressure dropped to 580.2~mbar, practically 10~mbar below the site median, 
with a temperature of $-0.1^{\circ}$C and relative humidity of 92\%. The same day 
registered the lowest daytime pressure, 580.8~mbar, at 10am, when the temperature 
had {\em dropped} to $-0.3^{\circ}$C. Bad weather occasionally occurs early in
the year, like on the 17/1/2004 at 6:20~am when the pressure reached its lowest 
dry season value, 581.6~mbar, with a temperature of $-3.7^{\circ}$C and 85\%
relative humidity.

As shown in Fig.~\ref{pres3dq2}, there is no significative difference between 
the cumulative distribution of atmospheric pressure values between day and night,
presumably because of the actual 12~hour cycle. The seasonal distributions
show that pressure tends to be 0.57~mbar lower during the dry season compared
to the wet season.
The grey scale diagram in Fig.~\ref{pres3dq2} shows that the main seasonal
effect on pressure  seems to be a  shift the daily cycle to later hours in June - July;
the minima move from 4~am and 4~pm in December/January to 6~am/6~pm in June/July.

\begin{figure*}
\includegraphics[width=\columnwidth]{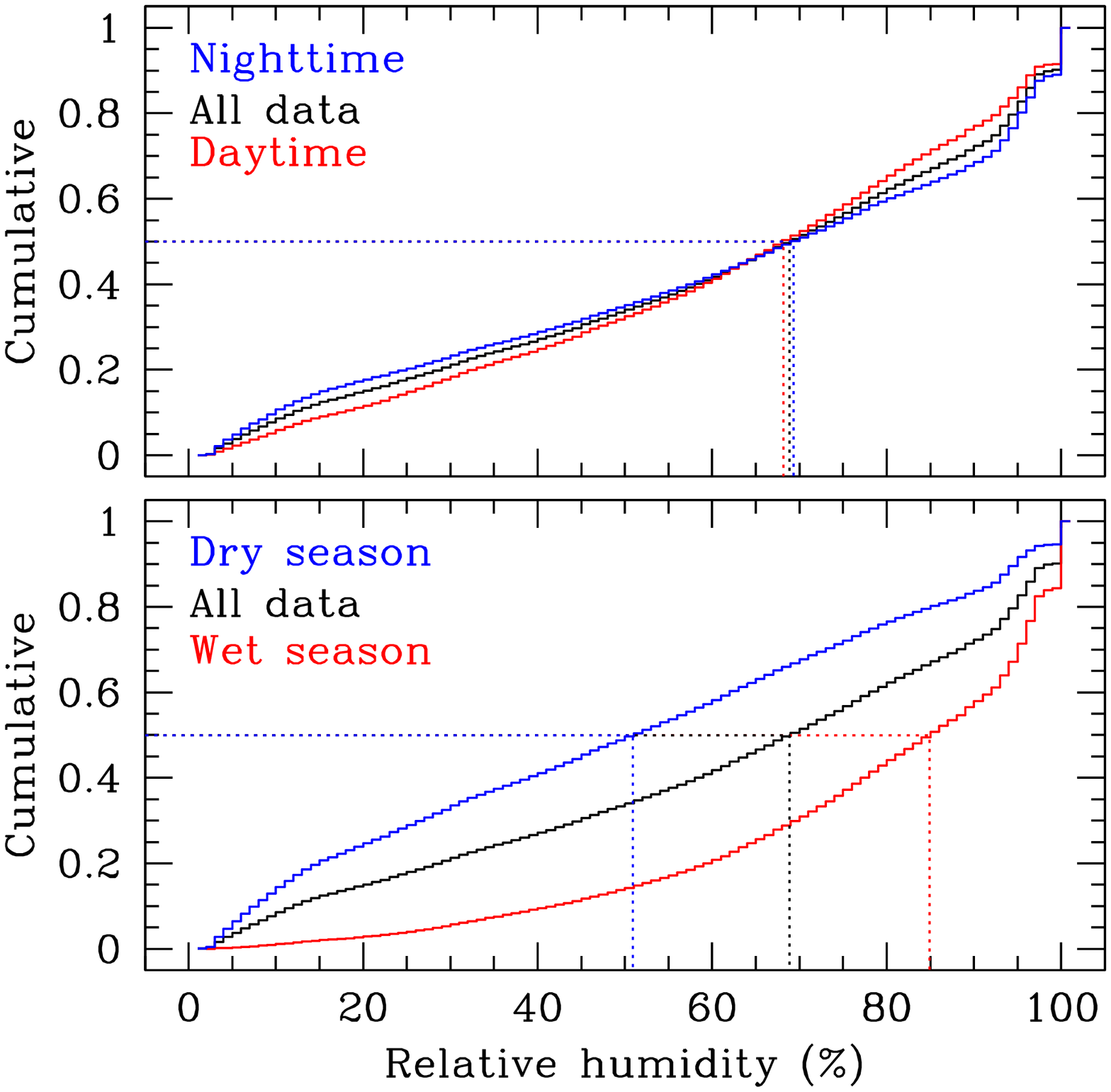}
\includegraphics[width=\columnwidth]{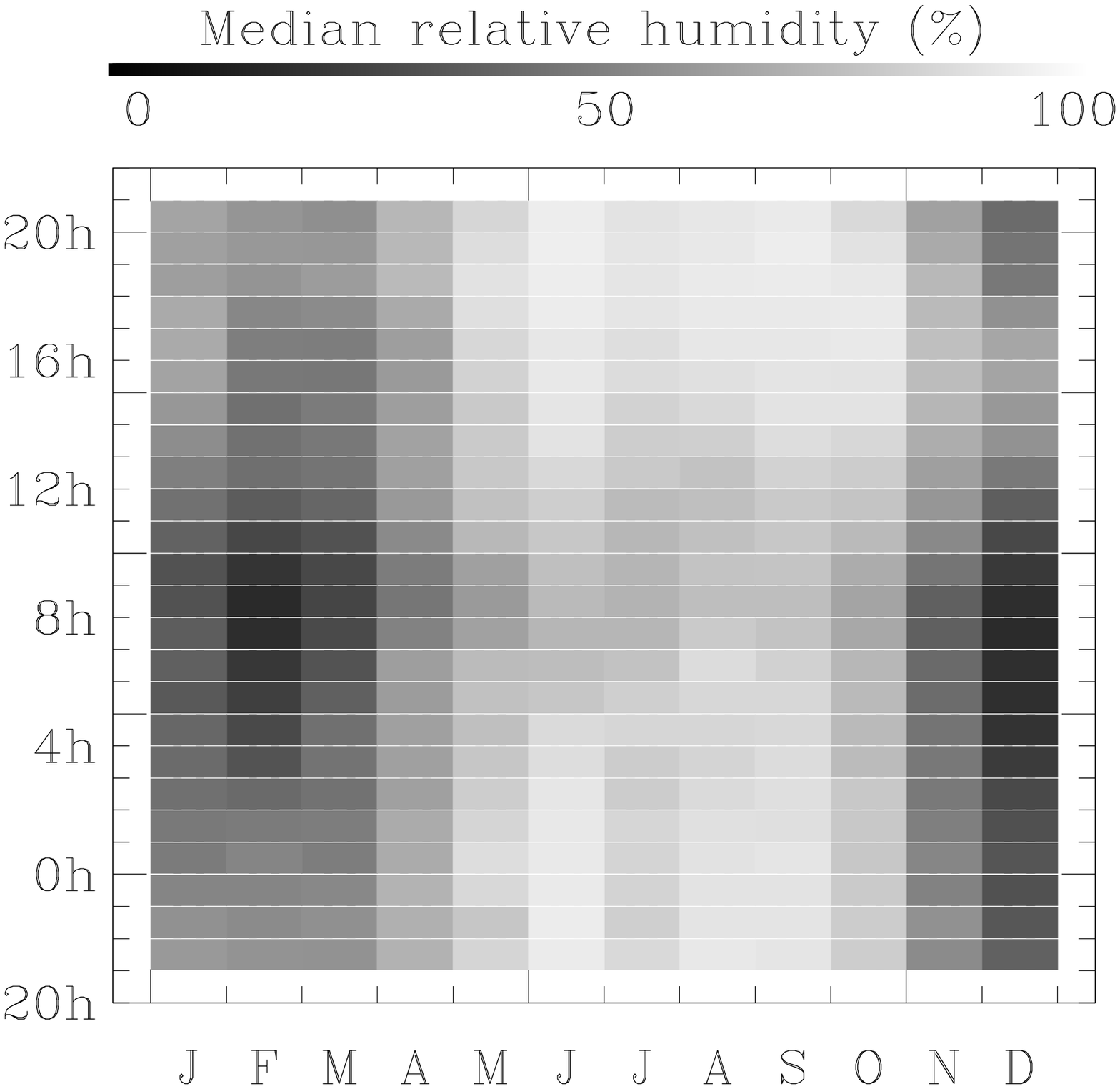}
\caption{Top left: the relative humidity distribution for the whole data set, for nighttime 
and  for daytime have been included only for completness as the RH is strongly seasonal dependent. 
 Bottom left:  distributions for the dry and the wet seasons:  the RH median  during the dry 
season is 50.92\%  while during the wet season is  84.92\%.  Right: a three dimensional representation
 of the relative humidity behaviour where the intensity corresponds to the RH median value 
for a given month and hour. The seasonal differences are clearly appreciated. 
See the electronic edition of MNRAS  for a color version of this figure. }\label{hum3dq1} 
\end{figure*}

\begin{figure*}
\includegraphics[width=\columnwidth]{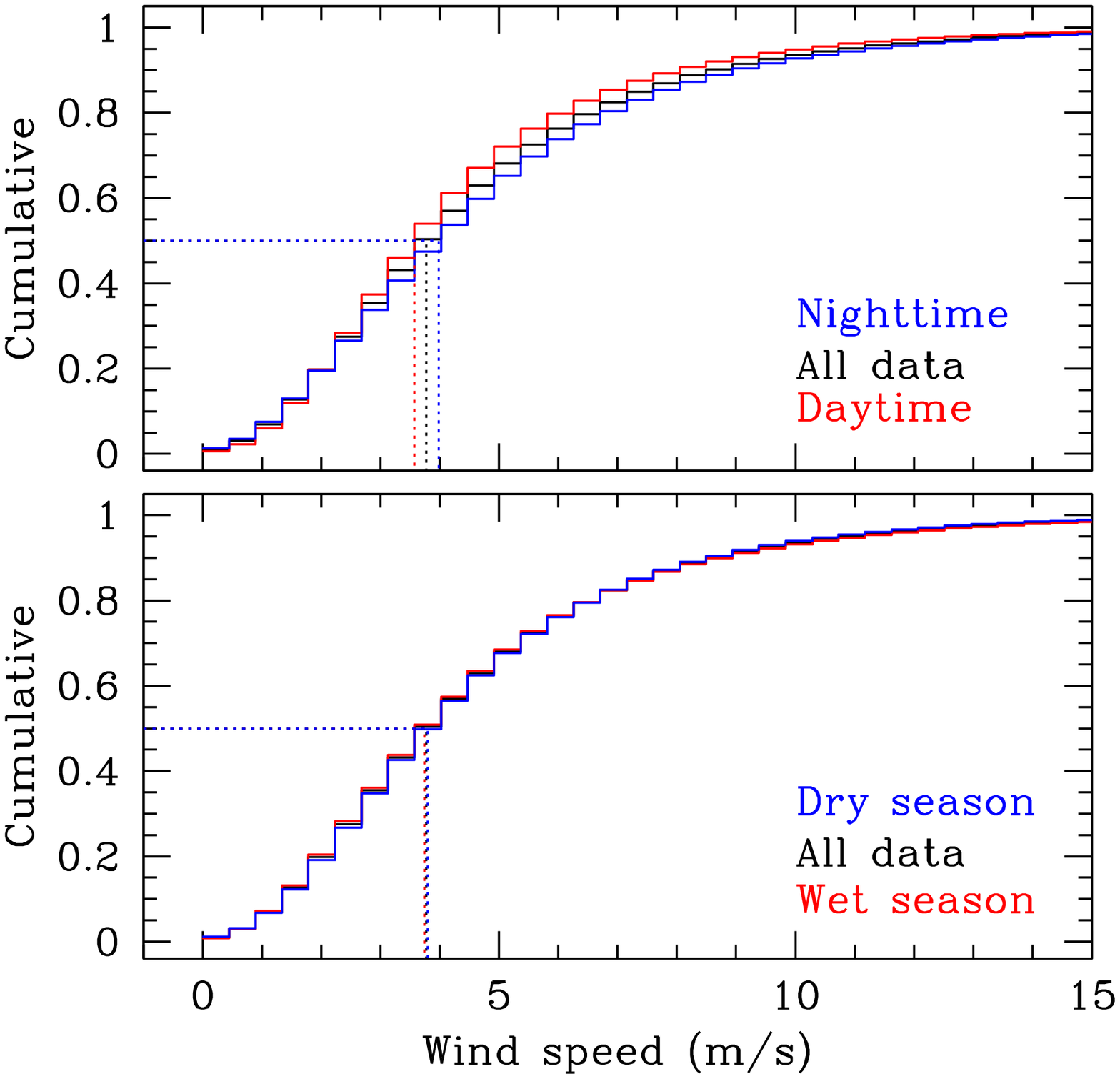}
\includegraphics[width=\columnwidth]{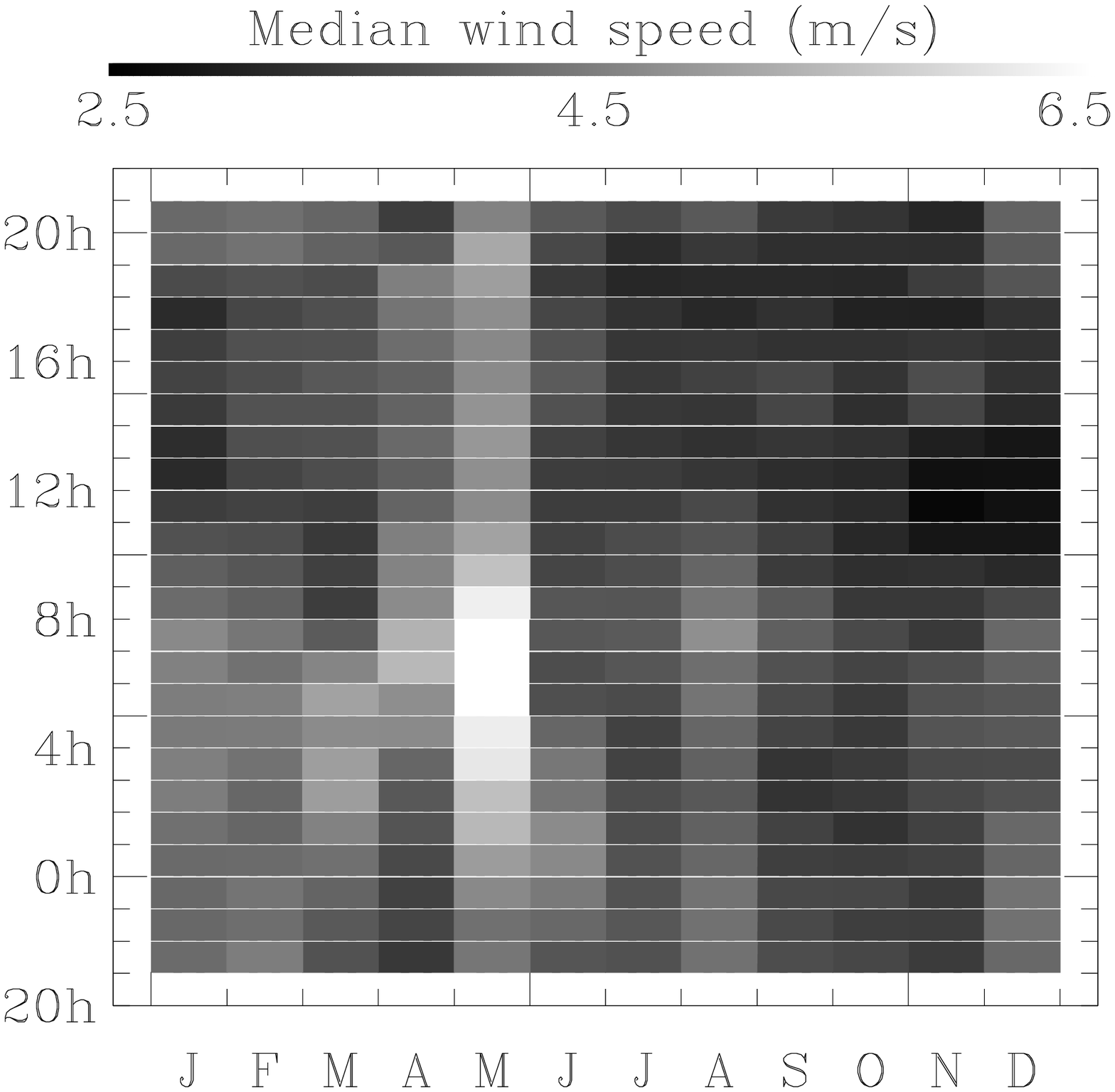}
\caption{Right: The wind  velocity distribution for  the whole data set is shown in 
black with a median  of 3.77~m~s$^{-1}$.  The distributions for  nighttime, daytime 
dry  and wet seasons are shown as indicated. The wind  velocity  median
 values are: at nighttime  3.98~m~s$^{-1}$,  at daytime 3.57~m~s$^{-1}$,  in the
dry season 3.80~m~s$^{-1}$ and  in wet season 3.74~m~s$^{-1}$. Left: a three 
dimensional representation of the wind speed, the intensity corresponds to the
wind velocity median for a given month and hour. The winds are stronger during
the nights. See the electronic edition of MNRAS 
for a color version of this figure. }\label{viento3dq2} 
\end{figure*}

The temperature and atmospheric pressure agree with a standard atmosphere model,
\begin{equation} 
 T(z) = T_{0} - \theta{z},  \quad P(z)= P_{0}(1 - \theta{z}/T_{0})^{\alpha}, 
\end{equation}
with the usual temperature gradient of a standard atmosphere, 
$\theta = -dT/dz = 6.5^{\circ}\rm C\, km^{-1}$, and the constant $\alpha = 
\mu m_{H}g/k\theta \simeq 5.256$, with $m_{H}$ the atomic mass unit, $g$ 
the acceleration of gravity, $k$ the Boltzmann constant, $P_{0}=1013.25~{\rm
mbar}$ and $\mu = 28.9644$ is the mean atomic mass of air. The data departure 
from the standard model requires a warmer base temperature, 
$T_{0}= 31^{\circ}{\rm C}  \simeq 304 K$, which results in $T(4.6~{\rm km}) 
= 1.1^{\circ}{\rm C}$ and $P(4.6~{\rm km}) = 588.2~{\rm mbar}$, close to 
the measured value. A warm standard atmosphere model appears reasonable for
the site although  it would be convenient to validate it with measurements of pressure
and temperature at different elevations.

\subsection{Relative humidity}

 The median RH  is 68.87\%  with quartile values of 36.76\%  and 92.59\%. The RH 
 values for day and nighttime are 68.19\% and 69.35\% . 
When folded by months, the data show a clear seasonal modulation with lower
values between November and March, $\la 50\%$ and higher humidity
between June and October with a median $\sim 90\%$  as illustrated
in Fig.~\ref{monthly_hour}.  A second clear trend is an increase of the RH at around 8h 
to reach a maximum at 18h, $\sim$80 \%. Once the Sun sets the RH starts decreasing to 
reach its minimum  value  of 49\%. Nevertheless, it  must be mentioned that for 
the daily cycle plot  the very low RH values measured  during the dry season are
merged with those obtained during the wet season at the same hour. 

   The cumulative  distributions of RH  are shown on the left side of Fig.  \ref{hum3dq1} where
the seasonal differences are better appreciated. For the
dry season the first, second and third quartile are 20.82\% 50.92\% and 78.52\%. In
contrast for the wet season the corresponding values are 64.86\%, 84.92\% and 96.18\%.
The right hand side of the same figure shows the median per hour and month in a grey 
scale. For November, December, January and Februrary, the driest times are from about 8:pm  up
to noon while for February, March, April and May the RH is lowest  from dawn up
to midday.  

\subsection{Wind velocity}
Wind velocity is an important factor for the Large Millimeter Telescope, specified to
perform at $\lambda\leq 1~\rm mm$ for wind velocities below 9~m~s$^{-1}$. Both stations 
give similar percentage of data below the critical value of 9~m~s$^{-1}$ 
(Davis: 91.5\%; Texas: 87.7\%). The Davis weather station has two wind values 
in each data record: one (``wind") corresponding 
to a mean  value acquired during the sampling interval ($\geq 1~\rm minute$) and a second
one (``whigh") corresponding to the maximum value during the same time interval. The median
value of whigh is 6.03~m~s$^{-1}$ and whigh$\geq 9~\rm m~s^{-1}$ for 22\% of the time.

The wind is fairly constant at the site, with a mild decrease less than  $ 1~\rm m~s^{-1}$
during daytime compared to nighttime. Differences between months are also small, except for
a marked increase in wind velocities during the month of May ($\sim 1.5~\rm m~s^{-1}$), noted 
by both datasets, and a small decrease ($\la 0.5~\rm m~s^{-1}$) of wind velocities in
the last months of the year.

The wind  distributions are shown in  Fig.~\ref{viento3dq2} in black for the 
whole data set; for nighttime, daytime, dry and wet seasons as marked.   In the 
3-D plot  a seasonal pattern can not be as clearly identified  as in the case of
other parameters but we can still notice that the wind is slightly  higher  during
the nights and the effect is more pronounced  during  the winter months. A special 
mention deserves the strong  winds in one year in May as can be seen from
 Fig.~\ref{monthly_hour}. The daily cycle  is better appreciated in the right panel of the
 same  figure  if we look at  the third quartile pattern.

The LMT has two other specified wind limits: operations at any wavelength are to stop 
if wind velocities reach 25~m~s$^{-1}$ and the telescope has to be stowed. In the extreme,
the design survival wind speed is   70\, m~s$^{-1}$ ( $\simeq$250\,\rm km/h). The two data sets
show extremely rare wind velocities above 25~m~s$^{-1}$, with whigh exceeding that value
0.3\% of the time. The largest wind speed registered so far corresponded to the storm
recorded on the 22$^{\rm nd}$ of February 2002, with $\rm whigh = 42.5~m~s^{-1}$. More
recently the near passage of hurricane Dean in August 2007 gave peak wind speeds of 
40.7~m~s$^{-1}$, the highest endured successfully by the Large Millimeter Telescope 
prior to the installation of its stow pins.

\section{Solar radiation and inferred cloud coverage}
\label{radiacion}

Solar radiation data was acquired by the Texas weather station between 
April 25, 2002 and March 13, 2008. The coverage for this time interval was  
62\%  (Table~3), with due consideration of the diurnal cycle. The data
are output as time ordered energy fluxes in units of W/m$^{2}$.
We obtain daily plots of the radiation flux which show the expected Solar 
cosine modulation. We present here a preliminar
analysis regarding a method to retrieve the cloud coverage from the
radiation data.

The radiation flux at ground level is modulated by the position of the Sun 
according to
\begin{equation}
F(t) = F_{\sun}\cos\theta_{\sun}(t)\,\psi(t)\, ,
\end{equation}
where $F_{\sun}$ is the solar constant, which for working 
purposes we take as exactly equal to $1367\,\rm W/m^{2}$; $\theta_{\sun}$ 
is the zenith angle of the Sun and $\psi(t)$ is a time dependent factor, nominally 
below unity, which accounts for the instrumental response, the atmospheric 
absorption on site and the effects of the cloud  coverage on the radiation transfer
through the atmosphere.

\begin{figure}
\includegraphics[width=\columnwidth]{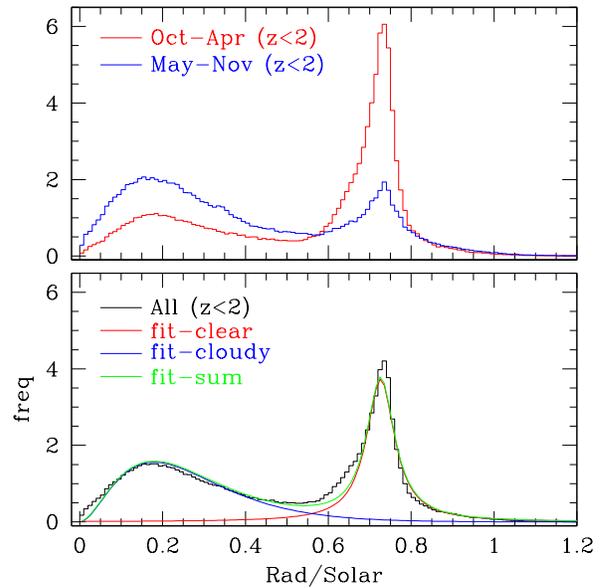} 
 \caption{{\em Top:} the observed distribution of
normalized solar fluxes for the wet and dry seasons.
{\em Bottom}: the bimodal distribution of the solar flux divided by the nominal 
solar flux at the top of the atmosphere, $F_{\sun}\cos\theta_{\sun}(t)$.
The distribution shows a bimodal behaviour which can be reproduced by a two 
component fit, show in solid lines. The relative area of both components
determines the clear/cloud fraction. See the electronic edition of MNRAS 
for a color version of this figure. \label{rad-histfit}} 
\end{figure}

Given the site coordinates, we computed the modulation factor $\cos\theta_{\sun}$ 
as a function of day and local time. Local transit cosine values range between 1 
around May 18 and July 28 (for 2008) and 0.74 at winter solstice (December 21).
Knowing the position of the Sun at the site as a function of time, we can study
the variable $\psi = F/(F_{\sun}\cos\theta_{\sun}) . $ 
The histogram of values of $\psi$ is shown in Fig.~\ref{rad-histfit}. It has a 
bimodal distribution, with a first maximum at around $\psi\sim 0.2$ and a narrow 
peak at $\psi\sim 0.75$, with a minimum around 0.55. 
We interpret the narrow component as due to direct
sunshine, while the broad component is originated when solar radiation is partially 
absorbed by clouds; we then use the relative ratio of these as the ``clear weather
fraction". Separating the data in intervals of $\cos\theta_{\sun}$, we observe that
the minimum of the distribution of values of $\psi$ increases with  $\cos\theta_{\sun}$
for small airmasses to become constant at lower Solar elevations, 
following the empirical relation:
$$\psi_{min} = \left\{
\begin{array}{lcl}
0.44\cos\theta_{\sun}+0.195 & {\rm for} & \cos\theta_{\sun}>0.625\, ,\\
0.47 & {\rm for} & \cos\theta_{\sun}\leq 0.625\, .\end{array} \right.
$$
For this first  analysis we separated data 
with $\psi\leq \psi_{min}$ as cloudy weather and data with $\psi>\psi_{min}$ as clear
weather. We computed the fraction of clear weather (clear/clear+cloudy), 
for every hour of data. Only hours with at least 30 minutes of data were
considered for the analysis, adding to 15223 hours of data. We considered
data with airmasses lower than 10.

\begin{figure}
\includegraphics[width=\columnwidth]{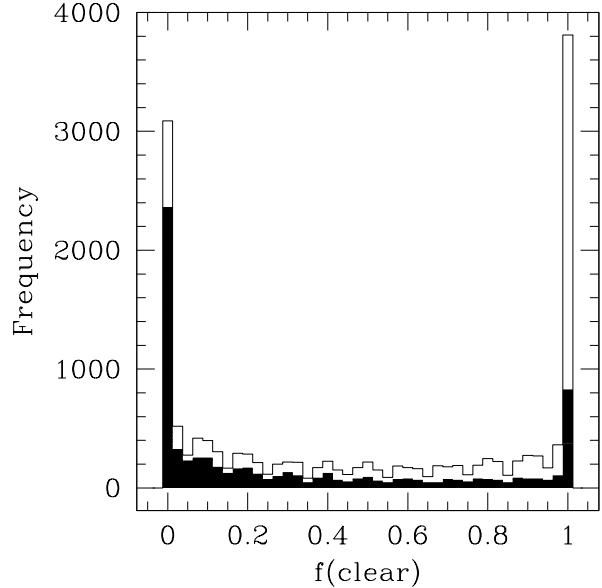}
\caption{Distribution of hourly clear fraction for the 15223 datapoints available.
The dark histogram shows the distribution for 7267 points of data taken in 
the (wet) months from May to October. The dry months are represented by
the difference between both histograms.\label{histograma}}
\end{figure}

The median clear fraction for the site is 48.4\%, consistent with values reported
by~\citet{Erasmus02}. In a  comprehensive study for the California Extremely 
Large Telescope (CELT) project, the authors surveyed cloud cover and water 
vapor conditions for different  sites  using  observations from the 
International Satellite Cloud  Climatology Project (ISCCP).  The study 
period is of  58  months between  July 1993 to December 1999 using a 
methodology that had been  tested and successfully  applied in previous studies. 
For Sierra Negra they measured a clear fraction for nighttime  of 47\%. 

We note that the set of hourly clear fractions behaves in a rather bimodal
fashion, as show in the histogram in Fig.~\ref{histograma}:  20.3\% of the 
hours have $f({\rm clear})=0$, while 25.0\%  have $f({\rm clear})=1$. The 
remaining  (55\%) have intermediate values. The histogram has a strong
modulation in terms of wet and dry months. If we consider the semester
between May and October the $f({\rm clear})=0$ peak contains 
32.5\% of the data, while the  $f({\rm clear})=1$ has 11.4\%. During the 
complementary dry months the $f({\rm clear})=0$ peak contains
9.0\% of the data, while the  $f({\rm clear})=1$ has 37.4\%. Intermediate 
conditions prevail around 55\% of the time in both semesters.

\begin{figure}
\includegraphics[width=\columnwidth]{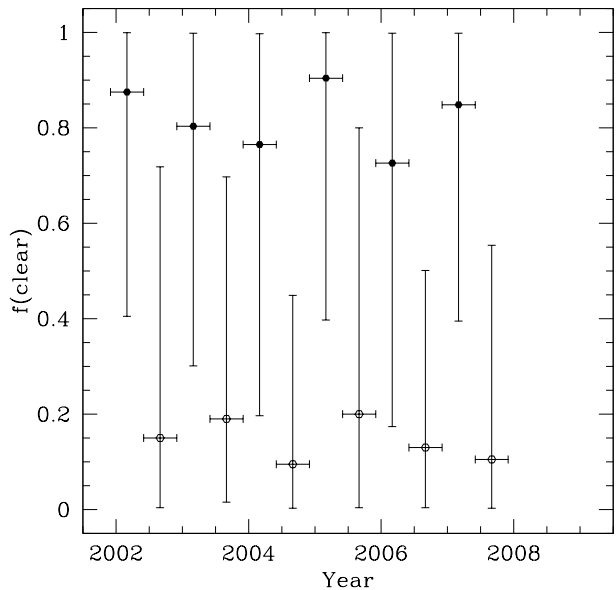}
\caption{Graph showing clear fractions for the different season. Points are 
at median; bars go from 1st to 3rd quartile. Wet season (open dots) is the 
yearly interval from May to October; dry season (full dots) is from November 
to April of the following year. \label{wetdry}}
\end{figure}

\begin{figure}
\includegraphics[width=\columnwidth]{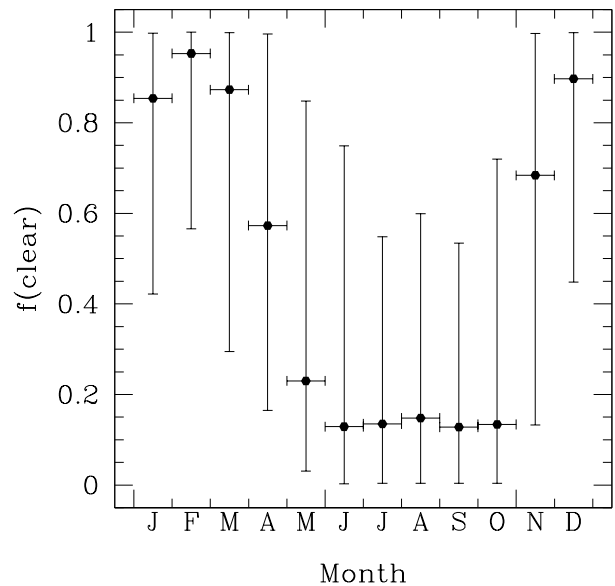}
\caption{Graph showing the median and quartile values of the fraction of clear
weather for the different months of the year.
\label{cloud_meses}}
\end{figure}

The contrast between dry and wet semesters is well illustrated in 
Fig.~\ref{wetdry}, showing the median and quartile fractions of clear time for
successive wet and dry semesters. Semesters are taken continuously, from 
May to October representing the wet season and November to April of the
following year for the dry season. The bars represent the dispersion in the
data, measured by the interquartile range. Large fluctuations are observed 
at any time of the year. The
contrast between the clearer dry months, with median daily clear fractions 
typically above 75\%,  and the cloudier wet months, with median clear fractions
below 20\%, is evident. The seasonal variation can be seen with more detail
in the monthly distribution of the clear weather fraction, combining the data 
of different years for the same month, shown in Fig.~\ref{cloud_meses}. 
The skies are clear (f(clear)$>80\%$) between December 
and March, fair in April and November (f(clear)$\sim 60\%$), and poor 
between May and October (f(clear)$<30\%$). 
The fluctuations in the data are such that clear fractions 
above 55\% can be found 25\% of the time
in the worst observing months.  

\begin{figure}
\includegraphics[width=\columnwidth]{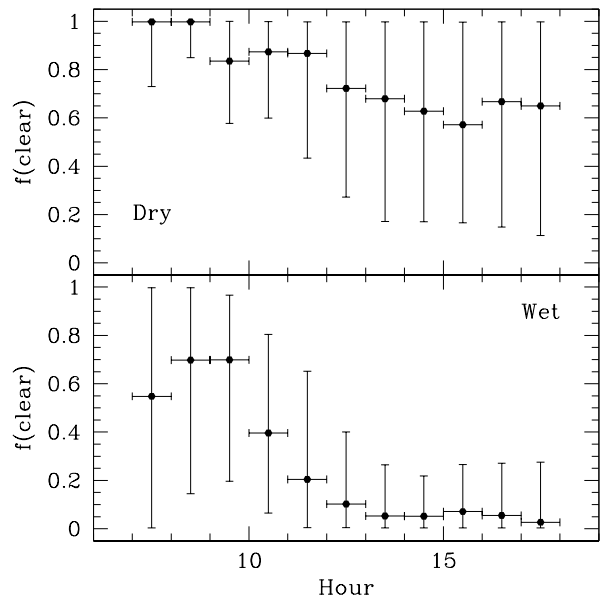}
\caption{Graph showing the median and quartile values of the fraction of clear
weather for each hour of day. The lower
and upper are for wet (MJJASO) and dry (NDJFMA) semesters respectively.
\label{cloud_horas2}}
\end{figure}

Fig. \ref{cloud_horas2} shows the median and quartile clear fractions as
function of hour of day for the  wet/dry subsets. The interquartile range practically 
covers the (0-1) interval at most times. We note that good conditions are more
common in the mornings of the dry semesters, while the worst conditions 
prevail in the afternoon of the wet season, dominated by Monsoon rain
storms. The trend in our results  for daytime is consistent with that obtained by
Erasmus and Van Staden (2002). By analysing  the clear fraction during 
day and nighttime they found that the clear fraction is highest before noon, 
has a minimum  in the afternoon and increases during nighttime.

\begin{figure}
\includegraphics[width=\columnwidth]{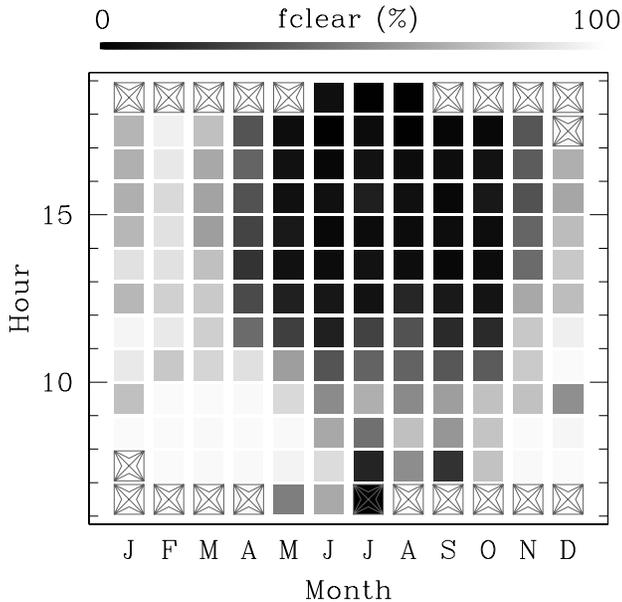}
\caption{Grey level plot showing the median fraction of clear time for
each month and hour of day. Squares are drawn when more than 10 
hours of data are available; crosses indicate less than 30 hours of data.
\label{cloud-msh}}
\end{figure}

Fig.~\ref{cloud-msh} shows a grey level plot of the median percentage 
of clear time for a given combination of month and hour of day. Dark
squares show cloudy weather, clearly dominant in the afternoons of the
rainy months (MJJASO). These are known to be 
the times of stormy weather in the near-equator. Clear conditions 
are present in the colder and drier months (NDJFMA). This plot is
similar to that of humidity. In fact, when relative humidity decreases, the fraction of
clear time  increases. The relation between  RH and $f({\rm clear})$ will be the subject
of a forthcoming paper.

\section{Summary and conclusions}
We have presented for the first time data and analysis of long-term meteorological
data directly obtained from local meteorological stations  at Sierra Negra.  A comparison
of the measurements from two weather stations  was carried out by cross calibrating 
the data; to include the accuracy errors of both stations, we obtained a fit for 
each parameter by  minimizing  $\chi^{2}$. In the case of the  temperature the 
values of both stations are consistent. For the wind velocities the fit is not
consistent wit the equality between the two data sets. However, we showed  
that their statistical behaviour is similar, probably the two stations are 
sampling the same wind but not simultaneously and the  differences might 
be due to the topography of the site. We will present a more detailed 
analysis of the wind in a forthcoming paper. 
The relative humidity sensor of one of the 
station slides up with time  providing data higher than the real ones. We 
verified our results with a  third station and data from the NCEP/NCAR 
 Reanalysis project database.
In the case of atmospheric pressure and solar radiation we only have data  
from one of the  stations. We reported  the daily, seasonal and annual behaviour of
temperature, atmospheric pressure, relative humidity, wind speed and solar
radiation. The site presents a  median temperature of 1.07~$^\circ$C and
an atmospheric pressure median of 590.11~mb.  The results for these two parameters
agree with a warm standard atmosphere model for which the   base temperature would be 
 $T_{0}=30.9^{\circ}$C.  As the site  is influenced by the
tropical storms moving off the Gulf of Mexico the median relative humidity  has a strong
seasonal dependence:  while the median value for  dry season is  50.92\% 
for the wet season is 84.92\%. The wind velocity median  is  3.77~m~s$^{-1}$, with
a third quartile of 5.88~m~s$^{-1}$ and a maximum  of 36.2~m~s$^{-1}$; 
these  values are below the three  LMT specifications: to perform below 1 mm the wind
speed must  below 9~m~s$^{-1}$; operation at any wavelength are stop if the wind velocity
is 25~m~s$^{-1}$ and the design survival wind speed is 70~m~s$^{-1}$. 
From the solar  radiation data we  developed a model for the radiation that allowed us  
to  estimate the fraction of time when the sky is clear of  clouds. The results obtained  are consistent with  
Erasmus and Van Staden (2002)  measurements  of cloud cover using satellite data. This
consistency shows the  great potential of our method  as cloud cover is a crucial parameter
for astronomical characterization of  any site. To our  knowledge
this is the first time that solar radiation data from the ground are  used to  estimate the
temporal fraction of clear sky.
The result presented here show that the meteorological conditions at Sierra Negra
are stable daily and seasonally  and have been so for the  seven years 
measured. We consider that this period is representative of the climate at the site.
Therefore Sierra Negra offers exceptional conditions
for such a high altitude, specially during the dry season, and is an ideal
site for  millimeter and high energy observations.


\section*{Acknowledgements}
NCEP Reanalysis data provided by the NOAA/OAR/ESRL PSD, Boulder, Colorado, USA, 
from their website at http://www.cdc.noaa.gov/. The authors thank G. Djordovsky,
A. Walker, M. Schoeck and G. Sanders for their kind permission to use
the results  from the Erasmus and Van Staden (2002) report for Sierra Negra.  
 Remy Avila and Esperanza  Carrasco thanks CONACyT support through the grant No. 58291.


\end{document}